\documentclass[journal=jpcafh, manuscript=article, layout=twocolumn]{achemso}

\usepackage{amsmath, amssymb}%
\usepackage{subcaption}%
\usepackage{multirow}
\usepackage[version=4]{mhchem}%

\usepackage{xr}%

\makeatletter
\newcommand*{\addFileDependency}[1]{
  \typeout{(#1)}
  \@addtofilelist{#1}
  \IfFileExists{#1}{}{\typeout{No file #1.}}
}
\makeatother

\graphicspath{}

\usepackage{color}

\title{From Spheres to Cones: Structural Instabilities and Acidity at Conical Regions 
  in Trivalent Metal Ion Nano-clusters}

\author{Jihong Shi}
\affiliation{
Department of Chemistry, The University of Western Ontario, London, Ontario, Canada N6A 5B7
}

\author{Han Nguyen}
\affiliation{
Department of Chemistry, The University of Western Ontario, London, Ontario, Canada N6A 5B7
}

\author{Mateo Pescador Arboleda}
\affiliation{
Department of Chemistry, The University of Western Ontario, London, Ontario, Canada N6A 5B7
}

\author{Styliani Consta}
\email{sconstas@uwo.ca}
\affiliation{
Department of Chemistry, The University of Western Ontario, London, Ontario, Canada N6A 5B7
}
\begin{document}


\begin{abstract}
Sub-nanometer aqueous clusters containing a single trivalent metal cation 
can exhibit charge-induced structural instabilities. 
Here, we present computational evidence that clusters containing a single 
\ce{Fe^{3+}}, \ce{Lu^{3+}}, or \ce{La^{3+}} ion undergo continuous geometric 
transformations as a consequence of this instability. 
These clusters dynamically evolve across their potential energy landscape, 
adopting triangular, elongated two-point, single-point, 
and more spherical configurations often with distinct conical surface protrusions.
The manifestation of this instability differs from that observed in mesoscopic and microscopic 
droplets containing macroions, where stable ``star-like'' structures form, 
characterized by a specific number of conical protrusions that varies with the droplet size. 
In the present study, we find that the orientation of the \ce{H2O} 
molecules surrounding the metal ion is influenced not only by the electric field of the trivalent ion
but also by the local conical protrusions. 
To further investigate the local acidity in the conical protrusions, 
we employ a proxy model system consisting of an aqueous nano-cluster containing 
three \ce{H3O+} ions, simulated using ab initio molecular dynamics. 
Within the conical regions of the cluster, protons exhibit diffusion 
across several water molecules, in contrast to the more localized proton 
delocalization observed in the compact body of the cluster. 
These findings suggest that local geometry can significantly modulate acidity
in highly charged nano-clusters, with potential implications for
understanding charge-transfer and ionization mechanisms in techniques such as
electrospray ionization mass spectrometry. 
Additionally, the structural motifs and solvent organization reported here provide a 
molecular-level framework that can complement interpretations from infrared spectroscopic data.
\end{abstract}

\section{Introduction}

Universality in the manifestation of physico-chemical properties across system 
sizes--from sub-nanometer clusters to microscopic droplets--is unusual.
For instance, phenomena such as melting and freezing \cite{berry1989freezing, de2023melting}, chemical reactivity \cite{signorell2022photoionization, gilligan2003solvation}, and ion solvation exhibit a strong dependence on system size \cite{burnham2006properties, stuart1996effects}.
In this work, we investigate charge-induced instability in clusters containing a single trivalent metal ion, exploring how this instability varies with both ion type and cluster size, and examining its implications for infrared spectral features.

A fundamental model due to Lord Rayleigh predicts the conditions of stability of a charged fluctuating droplet arising from its energy, which is expressed as the sum of the electrostatic energy and surface energy\cite{rayleigh1882, hendricks1963, oh2017droplets}. A linear stability analysis of the model leads to the Rayleigh limit\cite{rayleigh1882, hendricks1963}.
Mathematically, the Rayleigh limit of a conducting droplet is defined via the dimensionless Rayleigh fissility parameter, $X$, expressed as
\begin{equation}
  \label{eq:fissility}
  X = \frac{ Q^2} {64 \pi ^2 \gamma \varepsilon _0 R^3}
\end{equation}
where $Q$ is the droplet charge, $\gamma$ the surface tension, $\varepsilon_0$ and $R$ are the permittivity of vacuum and the radius of the droplet, respectively. 
When  $X = 1$ (Eq.~\ref{eq:fissility}) in a conducting droplet the droplet is at the Rayleigh limit. 
A droplet's Rayleigh limit is defined at the point where the surface tension forces that tend to compact a droplet are counterbalanced by the repulsive Coulomb forces among the unbalanced charges of
the same sign. 

A distinction between the stability of a conducting 
droplet\cite{rayleigh1882, hendricks1963, oh2017droplets} and of a droplet containing a single ion in a dielectric medium\cite{oh2017droplets} is discussed now. In a conducting droplet, the charge
carriers are many separate ions and the Rayleigh limit is given by $X = 1$ in Eq.~\ref{eq:fissility}. When the stability of a droplet containing a single ion, such as a lanthanide series trivalent ion 
or an ionized nucleic acid, in a dielectric medium is studied then  
$X$ is determined by Eq.~7 in Ref.~\cite{oh2017droplets}. For a certain value of the 
dielectric constant of the droplet, Eq.~7  yields the minimum value of $X$  by varying the order, $l$ of the spherical harmonics (it is noted that spherical harmonics are used to expand the droplet's shape fluctuations). Then the value of $X$ is used in Eq.~\ref{eq:fissility} to estimate $R$.
Another related difference between a conducting droplet and a single ion in a fluctuating spherical dielectric medium is that 
in the former case the first mode to become unstable is the $l=2$, while in the latter case the mode depends on the dielectric constant of the droplet\cite{oh2017droplets}.
It is noted that the Rayleigh limit of a conducting droplet arises from the  
generalized Rayleigh model for a dielectric droplet presented in Ref.~\cite{oh2017droplets} 
when the droplet's dielectric constant tends to infinity.
In a conducting droplet, the instability manifests as Rayleigh jets that emit a train  
of charged sub-droplets\cite{duft2003rayleigh, consta2022atomistic} from a larger parent droplet, while in a dielectric droplet in the form of ``star-like''-shaped droplets\cite{consta2010manifestation}. 

Computational evidence has shown\cite{oh2017droplets, consta2010manifestation, 
sharawy2016characterization} that mesoscopic and microscopic droplets charged
with a single macroion (e.g. a DNA molecule or an ionized protein) obey the predictions of a generalized Rayleigh's continuum 
model\cite{rayleigh1882, hendricks1963}
for
the onset of the charge-induced instability. 
Computations have also shown that at the instability regime the spherical shapes transform into 
stable ``star-like'' 
morphologies\cite{oh2017droplets, consta2010manifestation, kim2017ions, sharawy2016characterization}, 
where the number of points in the stars 
increases sequentially from one at the onset 
of instability to several for system parameters deeper into the instability regime.  

The generalized Rayleigh’s model\cite{oh2017droplets} is based on a linear stability analysis for small
fluctuations of the spherical shape.
Sub-nanometer sized clusters differ from the larger droplets in that
their physical chemistry is entirely dominated by the free surface and the large relative shape 
fluctuations. 
Therefore, it is hypothesized that in sub-nanometer sized clusters the Rayleigh model predictions for the onset of instability and the stable shapes in the instability regime
may differ from those in larger droplets.

In our study, the examples of \ce{Fe^{3+}}, \ce{Lu^{3+}}, \ce{La^{3+}} in aqueous clusters of sizes containing a few tens to a few hundreds of \ce{H2O} molecules are examined. Trivalent ions are selected because it is expected that the manifestation of the instability will be more pronounced in these higher charged ions than in the monovalent and divalent ions. The specific ions are selected as paradigms to demonstrate the dependence of the phenomena on the ion size.

Recognizing the instability regime in clusters comprised a few tens to a few hundreds of \ce{H2O} molecules and a trivalent metal ion is of practical importance for a number of experiments that measure the infra-red (IR) spectra of \ce{H2O} molecules in these cluster sizes\cite{prell2011structural, o2012effects}. 
Experimental methods such as black body infrared radiative 
dissociation (BIRD), and ensemble infrared photodissociation (IRPD) may
probe the extent at which an ion's electric field influences the structure of the surrounding hydration shells\cite{yang2019deconstructing}. 
These methods can be applied to clusters comprised a single ion and a few tens of \ce{H2O} molecules
up to several hundreds \ce{H2O} molecules\cite{prell2011structural}.
Williams et al. demonstrated that the IR spectra of aqueous clusters comprised a single
monovalent, divalent and trivalent ion in 36~\ce{H2O} molecules, reflect the 
reorganization of the hydrogen-bond network induced by the ion’s electric field, 
with a red shift in the free-OH (found in the cluster surface) 
stretching band proportional to the ion’s charge\cite{prell2011structural}. 
The ion-induced patterning in the aqueous clusters provided evidence that the 
 effect of multivalent ions beyond the first hydration shell in the bulk solution is considerable. 
 This is in agreement with the simulation results presented here that show that indeed the electric
 field of the trivalent ion, regardless of its size,
 strongly influences its immediate two hydration shells. We further 
 show that the effect of the collective electric field from the ion and the other \ce{H2O} molecules in the third and farther hydration shells orients the \ce{H2O} molecules differently in clusters subject to instability relative to clusters outside the instability regime. The difference implies that the ion's electric field also influences the third and partially fourth hydration shell in clusters that
 are subject to instability. 
 As it will be shown, in clusters where the effect of the
 instability has ceased, the
 trivalent ion is located by approximately three hydration shells below the surface. 
 The orientation of the third hydration shell, which also includes surface \ce{H2O} molecules,
 is influenced differently by 
 \ce{Fe^{3+}}, \ce{Lu^{3+}}, \ce{La^{3+}}, depending on the ion's size.

The charge-induced instability shapes are characterized by conical protrusions on the cluster surface. 
For the first time, the acidity of the \ce{H2O} molecules
in the conical protrusions, and consequently, their ability for proton transfer, is examined here. To probe the chemical reactivity in these highly convex regions, we use a proxy simulation set-up, where an aqueous cluster charged
with three \ce{H3O+} ions is modeled by ab initio molecular dynamics (AIMD). 
It is found that in a conical protrusion that persists for a few picoseconds of AIMD simulation, 
there is rapid proton diffusion over several \ce{H2O} molecules, 
unlike proton delocalization over two or three \ce{H2O} molecules found in the 
compact body of the cluster. The proton diffusion in the extruded regions complements studies that have been performed on the proton transfer in bulk solution along \ce{H2O} chains\cite{decoursey2014philosophy,natzle1985recombination, huckel19283,de1806decomposition,cukierman2006tu,agmon1995grotthuss, agmon2016protons,geissler2001autoionization,brewer2001formation, hassanali2011recombination,hassanali2013proton, tuckerman1995ab, newton1971ab, markovitch2008special}.

Understanding trivalent metal ion-\ce{H2O} interactions is significant on its own merits.  
These interactions play an important role in solution chemistry as for example in liquid-liquid phase separation (LLPS) of proteins\cite{saha2022trivalent}. Nevertheless their role in LLPS is less studied than the water-alkali and alkaline earth ions.
Specific ordering of \ce{H2O} molecules or of other solvents induced by a multi-valent ion in clusters may be also important in catalysis. Metal ion-solvent clusters comprised up to approximately twenty solvent molecules
have been actively explored over several decades for their distinct hydration and catalytic properties\cite{beyer2007hydrated, bohme2005gas, van2023surface, gilligan2003solvation, yang2025periodic, carnegie2020microsolvation, miliordos2014elucidating, yang2019deconstructing, armentrout1989chemistry}.

Preparing clusters of trivalent metal ions of the desired size
in IR experiments
poses a challenge 
because the metal ion may undergo hydrolysis within the cluster 
or the metal ion is electrosprayed already hydrolyzed from the parent bulk solution.
Williams et al. have reported\cite{prell2011structural, bush2008reactivity}
that the experimental
conditions under which the metal ions in the lanthanide series  do not hydrolyze in nanodroplets
have been found, while \ce{Fe^{3+}} always been
detected in its hydrolyzed form possibly because it is already hydrolyzed
in the parent bulk solution. In the present study the molecular modeling 
of \ce{Fe^{3+}}-\ce{H2O} clusters does not allow for the hydrolysis reaction
but it assists along with \ce{Lu^{3+}} and \ce{La^{3+}} to establish a trend 
that shows the dependence of the phenomena on the ion size.

In our study the majority of the simulations are performed at the lowest possible temperature, 190~K, where the \ce{H2O} molecules are still mobile and the energy transfer to the rotational motion of the cluster does not freeze the internal motions of the molecules.
However, the experiments are performed at cryogenic temperatures ($\lessapprox 133$~K), where the metal ion clusters are likely to be in ice-like states\cite{cooper2016delayed}. The  manifestation of the 
 instability at a temperature where the \ce{H2O} dynamics is very slow is also presented in this study. However, the emergence of ice-like structures as a function of cluster size when the dynamics is very slow deserves its own study that we currently perform.

\section{Methodology}
Here we present the main points of the computational methods.
A detailed account of the methodology is found in Sec.~S1-S3 in SI.
Systematic studies of molecular dynamics (MD) of aqueous clusters 
charged with a single \ce{Fe^{3+}}, \ce{Lu^{3+}}, or \ce{La^{3+}} ion
were performed using the Nanoscale Molecular Dynamics\cite{namd} (NAMD) v~2.14
software and visualized by Visual Molecular Dynamics\cite{HUMP96} (VMD) v~1.9.4.
The clusters were comprised of $N_{\ce{H2O}}=32-242$ and
a single \ce{Fe^{3+}}, \ce{Lu^{3+}}, or \ce{La^{3+}} ion.
A few simulations were also performed with $N_{\ce{H2O}}=800$ and a 
single \ce{Fe^{3+}}, \ce{Lu^{3+}}, or \ce{La^{3+}} ion.
MD simulations were also performed for pristine aqueous
clusters with $N_{\ce{H2O}} = 70$ and 242 for using them as reference systems.
The water molecules were modelled by the TIP3P-CHARMM  model \cite{mackerell1998all}
and the ions by the parameters reported in Ref.\cite{won2012force}.
We note here that the parameters for \ce{Fe^{3+}} from Ref.\cite{won2012force} lead to
a first hydration shell with $\sim 4$~\ce{H2O} molecules surrounding the ion.
This is to be constrasted with the coordination number of six that is found
in experiments. Regardless of the low coordination number of \ce{Fe^{3+}},
the \ce{Fe^{3+}} model still shows that the phenomena observed here do not depend
on fine details of the force field. For \ce{Lu^{3+}}, or \ce{La^{3+}} the 
coordination number in the first hydration shell is in agreement with the
experiments.
The systems were simulated at temperature $T = 90$~K, 130~K, 190~K and 250~K.
Among the tested temperatures, $T=190$~K was selected to perform the majority of the simulations 
because at this temperature
the molecules in the clusters are still
diffusing within the simulation time.
At $T = 90$~K and 130~K the clusters
demonstrate deformations caused by the instability but there is no internal motion of the
\ce{H2O} molecules and the ion
within the simulation time.

To study the proton diffusion a cluster containing 64~\ce{H2O} molecules and
3\ce{H3O+} ions were also simulated using \textit{ab initio} Born-Oppenheimer
MD (BOMD) in CP2K v~9.1 software package \cite{kuhne2020cp2k}. 
The validation of the simulation protocol is
presented in Sec.~S4 in SI.
The system was simulated at $T=300$~K using a Nos\'{e}–Hoover thermostat. The protonated cluster
was placed in a large vacuum box
of dimensions $48 \times 48 \times 48$~{\AA}$^3$.
A vacuum shell of 3-\ce5 \AA\ around the clusters was considered to be a sufficient and reasonable range for AIMD simulations of non-periodic isolated systems\cite{marx2009ab}.
The Poisson solver wavelet\cite{genovese2006efficient,genovese2007efficient} was chosen to solve the Poisson equation to satisfy the non-PBC. This solver
is compared to the Martyna-Tuckerman solver\cite{martyna1999reciprocal} in Sec.~S5 in SI. 
It is noted that this is the first usage of BOMD
in capturing the cluster fragmentation and observing directly the motions of several 
transferable protons in a protonated cluster.

\section{Results and discussion}

\subsection{Charge-induced structural instability}
Defining the charge-induced instability in sub-nanometer clusters entails several challenges. 
Deviations from predictions of linear 
theories\cite{rayleigh1882, hendricks1963, oh2017droplets} for the
onset of the charge-induced instabilities may arise by the high electric fields
applied from the ion in only two-three hydration shells that are present in the sub-nanometer cluster 
sizes
as well as large shape fluctuations relative to
the spherical shape inherent in small clusters. 
In particular, smaller ions such as \ce{Fe^{3+}} can produce high fields that 
tightly orient nearby water molecules, limiting their polarizability 
and disrupting the expected linear dielectric response.
Eq.~7 in Ref.\cite{oh2017droplets}
and Eq.~1 in this manuscript come from a linear theory\cite{rayleigh1882, oh2017droplets}. 
Because of the lack of a non-linear theory,
we employ them to provide an estimate of the
cluster size where the instability may manifest. As we will discuss, the simulations
show agreement with the theoretical predictions.
Equation~7 in Ref.\cite{oh2017droplets} for a certain value of the dielectric
constant, $\varepsilon$,  of a dielectric droplet  containing a single (macro)ion, yields the minimum 
$X$ by varying the order of the spherical
harmonics, $l$. This value of $X$ is used in Eq.~1 to approximately compute the radius, $R$, 
of the droplet and from $R$ the number of \ce{H2O} molecules, $N_{\ce{H2O}}$.
Combinations of values of $\varepsilon$ that vary in the range of 10 to 82, 
with those of surface tension, $\gamma$ 
in the range of 52.3~mN/m (which is the $\gamma$ for TIP3P \ce{H2O} model\cite{vega2007}
at 300~K) to 10~mN/m yields a minimum number of $N_{\ce{H2O}}=92$ below which
the instability may manifest.  A combination of $\epsilon= 10$ and $\gamma = 10$~mN/m 
yields $N_{\ce{H2O}}=242$, which is the prediction 
for the maximum number of \ce{H2O} molecules
below which the instability may manifest.

\begin{figure}[htpb!]
    \centering
    \includegraphics[width=1.0\linewidth]{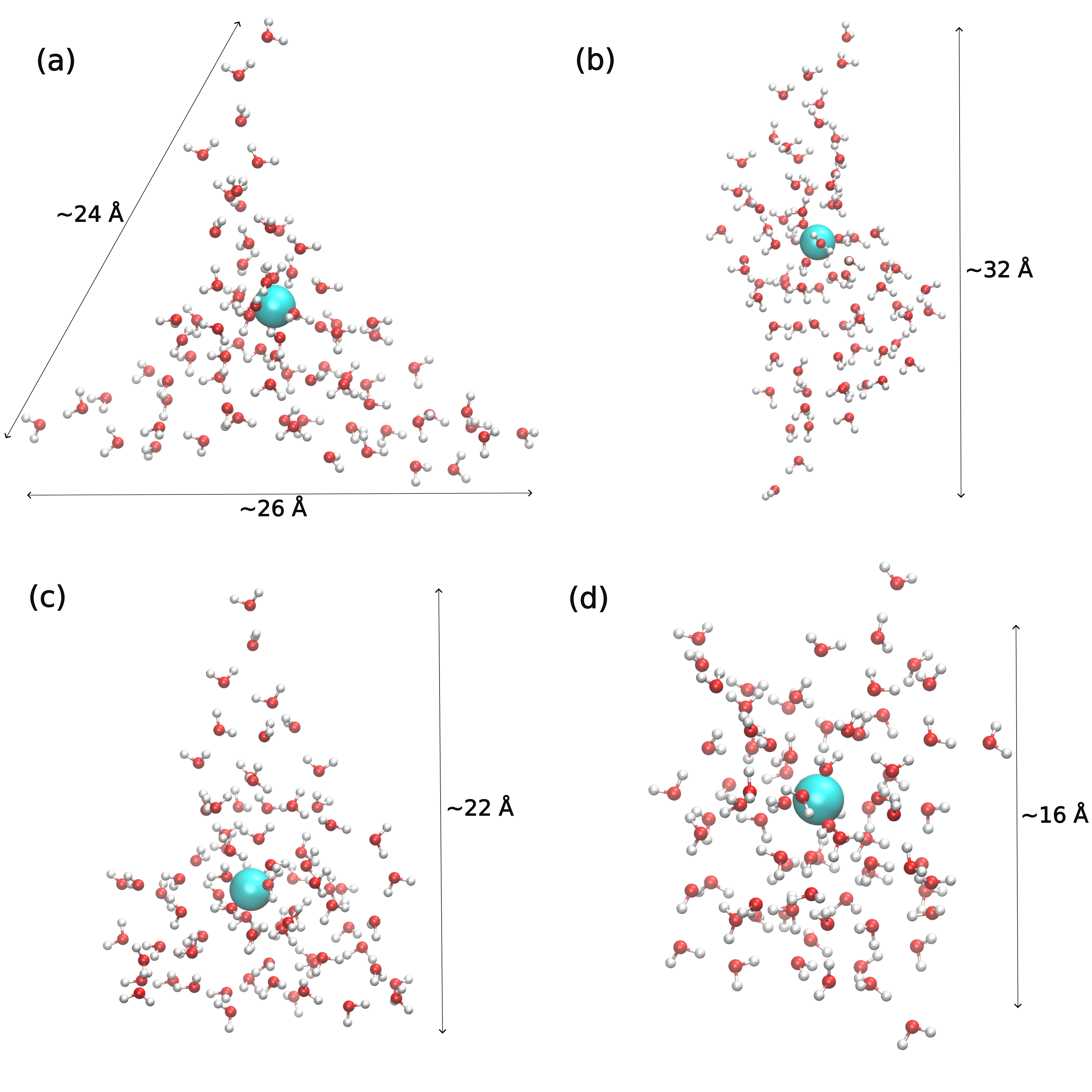}
    \caption{Typical snapshots of a cluster comprised a single \ce{Fe^{3+}} ion and 
    $N_{\ce{H2O}}= 78$ at 190~K over 40~ns production run reveal 
   four distinct motifs: (a) triangular, (b) elongated with two diametric conical tips, 
 (c) a single conical tip, and (d) a more spherical with a cleaved surface
 into several and shorter conical protrusions relative to the triangular shape. The color coding
 is: turquoise sphere represents the \ce{Fe^{3+}} ion, red oxygen sites, and white hydrogen sites.}
    \label{fig:78tip3-Fe3+-190K}
\end{figure}

\begin{figure}[htpb!]
    \centering
    \includegraphics[width=1.0\linewidth]{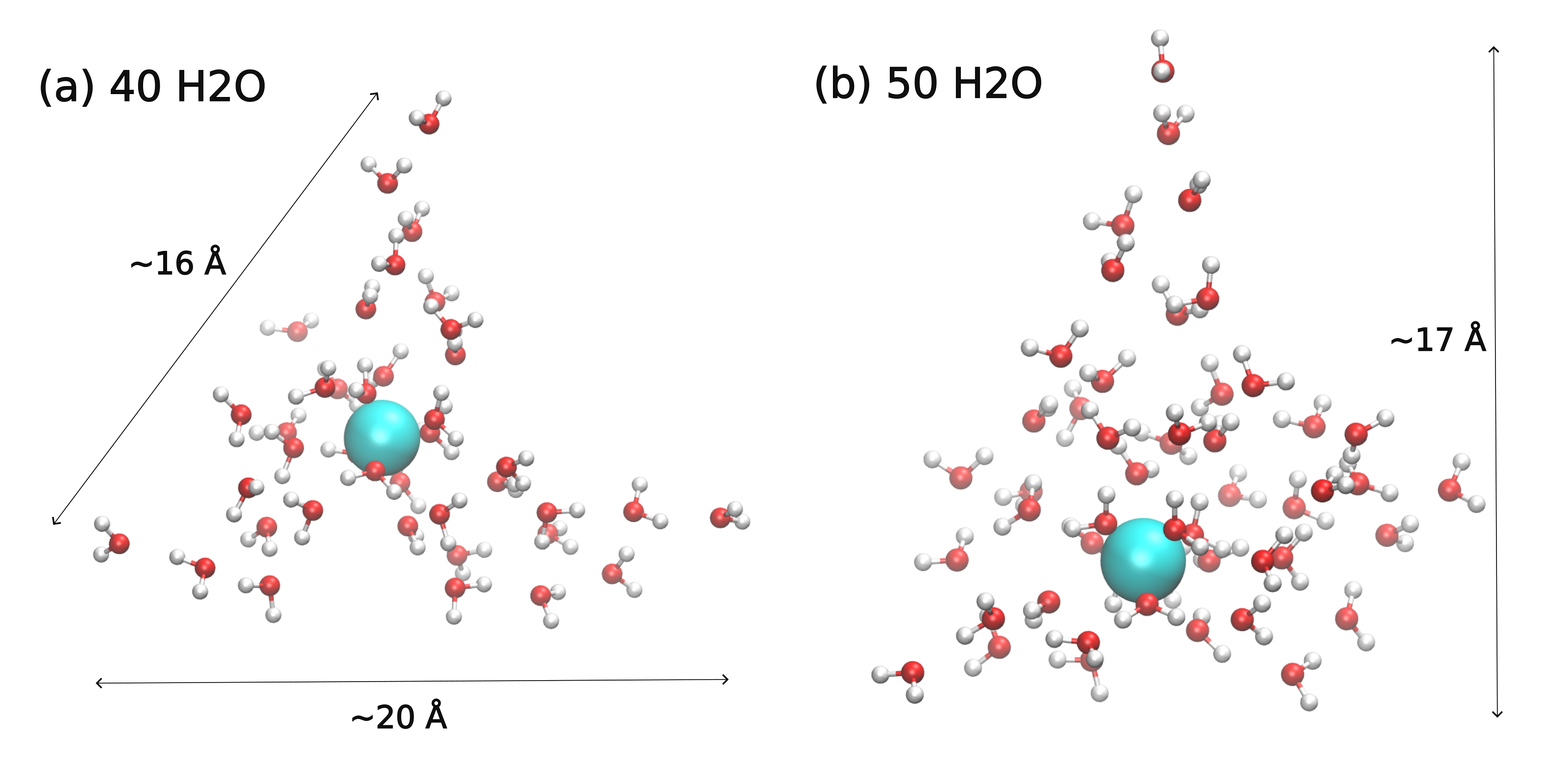}
    \caption{Typical snapshots of clusters comprised a single \ce{Fe^{3+}} ion and (a) $N_{\ce{H2O}}=40$, (b) $N_{\ce{H2O}}=50$ at 130~K. The color coding is the same as in 
    Fig.~\ref{fig:78tip3-Fe3+-190K}.}
    \label{fig:Ntip3-Fe3+-130K}
\end{figure}

The example of \ce{Fe^{3+}} is employed to demonstrate the instability and
is compared with \ce{La^{3+}}, and \ce{Lu^{3+}}.
Figure~\ref{fig:78tip3-Fe3+-190K} shows typical snapshots of cluster configurations for
a \ce{Fe^{3+}}-$N_{\ce{H2O}} = 78$ system at 190~K and a Movie in SI shows the
transitions. 
Within the energy landscape, the cluster dynamically evolves through configurations with a triangular 
(Fig.~\ref{fig:78tip3-Fe3+-190K}~(a)),  
elongated with two diametric conical tips (Fig.~\ref{fig:78tip3-Fe3+-190K}~(b)), 
one conical tip (Fig.~\ref{fig:78tip3-Fe3+-190K}~(c)), 
and a more spherical shape with 
several and shorter conical protrusions on the surface (Fig.~\ref{fig:78tip3-Fe3+-190K}~(d)) 
than the triangular shape. The continuous transition among the shapes is consistent 
with the fact the physical properties change continuously in clusters as for
example in phase transitions\cite{berry1989freezing, de2023melting}.
 These structures are to be compared 
 with the stable ``star-like'' shapes that we found\cite{consta2010manifestation,oh2017droplets} 
 in dielectric droplets 
 comprised $N_{\ce{H2O}} \gtrapprox 1000$ 
 and a single macroion such as a nucleic
 acid or a protein. In a droplet containing $N_{\ce{H2O}} \gtrapprox 1000$
 ``star-like'' shapes with a specific number of points are maintained for a range of droplet sizes as shown in Fig.~2 in Ref.~\cite{oh2017droplets}.
 As the droplet size changes 
 sharp transitions between ``star-like''-shapes with a different number of points were observed.  
 Occasionally, very near this transition point
 a droplet may visit ``star-like'' configurations with both $n$ and $n+1$ points.
 In contrast, in the sub-nanometer sized clusters, we find that
 ``star-like''-shapes with different number of points 
 are explored by the same cluster size at $T = 190$~K where the \ce{H2O} molecules
 are still mobile. These continuous transitions between different
 shapes may arise from the fact that the large
 relative shape fluctuations in the clusters lower the energy barrier among the various shapes.
Similarly to \ce{Fe^{3+}}, clusters at $T=190$~K comprised a single \ce{La^{3+}} or \ce{Lu^{3+}} and $N_{\ce{H2O}} < 100 $ show pronounced shape deformations
with one, two and three conical protrusions. Interestingly, the prediction of the analytical theory for the minimal number of \ce{H2O} molecules below which the structural instability may manifest is in agreement with the simulation results. For $N_{\ce{H2O}} > 100 $ the number of conical protrusions on the surface attenuate as the cluster size increases.

 At $T=130$~K, the \ce{H2O} molecules are immobile. Because of the
 suppression of the thermal fluctuations it may be expected that a specific ``star-like''-shape will appear 
 permanently within a long simulation. 
 Indeed, Fig.~\ref{fig:Ntip3-Fe3+-130K} for $N_{\ce{H2O}} = 40$ and 50, shows three-point (Fig.~\ref{fig:Ntip3-Fe3+-130K}~(a)) and one-point (Fig.~\ref{fig:Ntip3-Fe3+-130K}~(b))
 ``star-like'' shapes, respectively that are maintained throughout a simulation of 40~ns. 
 For $N_{\ce{H2O}} \gtrapprox 60$ the shapes do not show large conical deformations on the
 surface. The cease of the instability earlier than for droplets at 190~K is attributed to
 the fact that at 130~K the surface tension is higher and the slow dynamics may trap the cluster in certain long-lasting potential energy minima.
 The considerably 
 slow dynamics of the clusters at the low temperature requires a separate study that we currently perform
 to identify
 the emergence of ice-like structures and the relation to instability. 
 Aqueous clusters at 130~K composed of several tens of \ce{H2O} molecules and a single \ce{La^{3+}} or \ce{Lu^{3+}} ion also show one, two, three or more  distinct conical protrusions, similar to \ce{Fe^{3+}}.

\begin{figure}[htb!]
\centering
\begin{subfigure}[htbp]{0.5\textwidth}
\centering
\includegraphics[width=\textwidth]{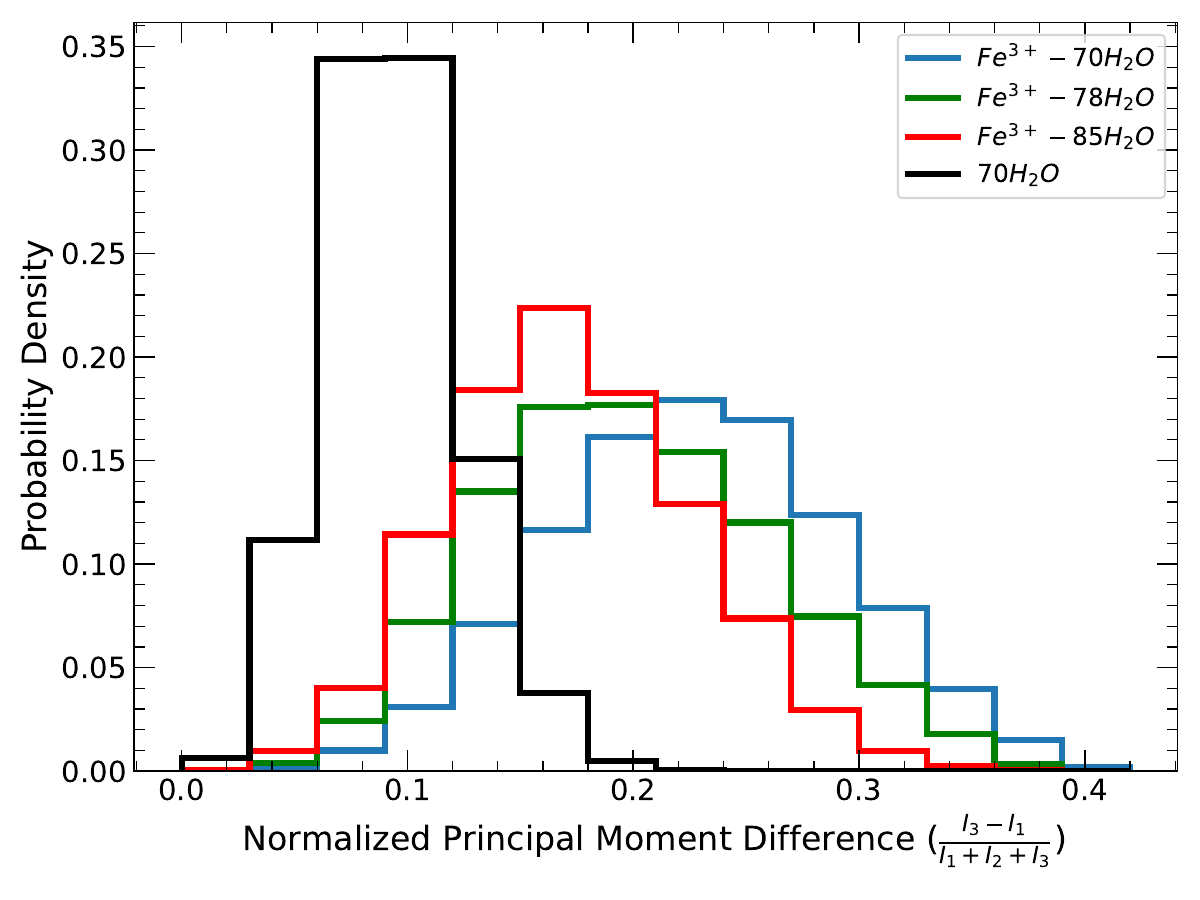}
\caption{}
\end{subfigure}
\begin{subfigure}[htbp]{0.5\textwidth}
\centering
\includegraphics[width=\textwidth]{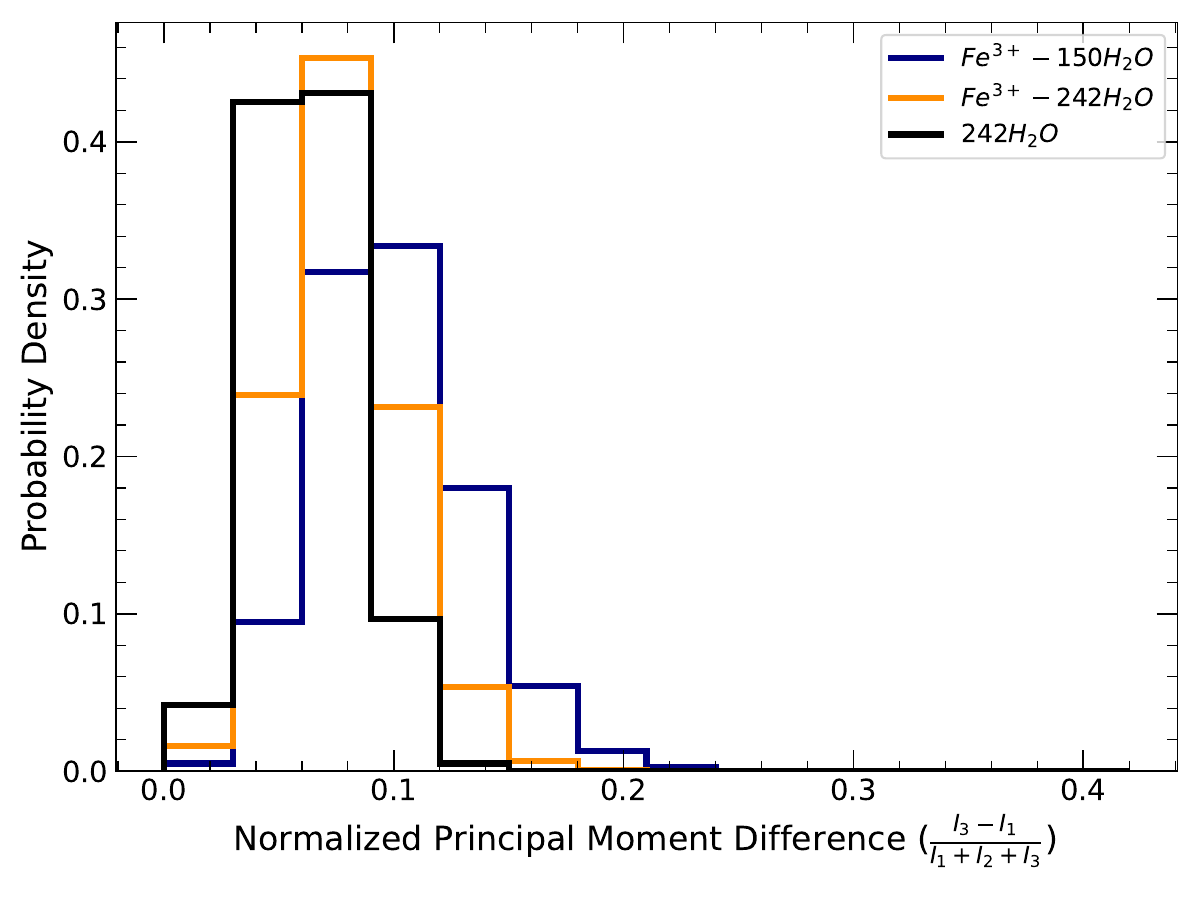}
\caption{}
\end{subfigure}
\caption{(a) Probability distribution of the ratio of the
      difference between the largest ($I_3$) and smallest ($I_1$) moments of inertia
divided by the sum of the
three moments of inertia ($I_1 + I_2 + I_3$) in \ce{Fe^{3+}}-$N_{\ce{H2O}} = 70-85$ clusters.
A pristine cluster comprised $N_{\ce{H2O}} = 70$ is used as a reference system.
The simulations were performed at $T= 190$~K and the production run was 40~ns. 
(b) Same as (a) but for
    \ce{Fe^{3+}}-$N_{\ce{H2O}} = 150-242$ clusters. The
    reference system is a pristine cluster comprised of $N_{\ce{H2O}} = 242$.
}
    \label{fig:moI-fe3+-190K-70h2o}
\end{figure}

We further investigate the distinct shape fluctuations
of the trivalent metal ion clusters 
in Fig.~\ref{fig:moI-fe3+-190K-70h2o}~(a) that shows the shape deviations of clusters 
comprised \ce{Fe^{3+}}-$N_{\ce{H2O}}=70-85$ from that of a pristine $N_{\ce{H2O}}=70$ cluster 
and, in Fig.~\ref{fig:moI-fe3+-190K-70h2o}~(b) the deviation of 
clusters comprised \ce{Fe^{3+}}-$N_{\ce{H2O}} = 150-242$ 
from a pristine cluster comprised $N_{\ce{H2O}} = 242$.
We define a descriptor
of the cluster shape by the difference between the largest ($I_3$) and smallest ($I_1$) moments of inertia 
divided by the sum of the
three moments of inertia, $I_1 + I_2 + I_3$. This descriptor nearly 
removes the effect of the cluster size from the comparison of the
shapes and as a result we can compare the shapes of different
cluster sizes shown in Fig.~\ref{fig:moI-fe3+-190K-70h2o}~(a) and (b). 

Figure~\ref{fig:moI-fe3+-190K-70h2o}~(a) and (b)  
show that shapes of clusters even with $N_{\ce{H2O}} \sim 150$ deviate from 
those of the pristine \ce{H2O} clusters. The $N_{\ce{H2O}} \sim 150$ do not
show well-defined conical protrusions as the clusters with $N_{\ce{H2O}} \lessapprox 100$ 
but they still show pronounced shape fluctuations because of the presence 
of the ion, that are different from those of a pristine aqueous cluster.
The analysis of the cluster shapes suggests that for $N_{\ce{H2O}} < 242$ the
orientation of the \ce{H2O} molecules will 
be influenced by the distinct 
geometries and shape fluctuations induced by the ion.

\subsection{Ordering of \ce{H2O} molecules} 
The range of the electric field of the ion shows its imprint in the structure of the
\ce{H2O} molecules surrounding the ions as it is described in the radial distribution
functions (RDFs) and in 
the orientation of the surrounding \ce{H2O} molecules. 

\begin{figure}[htbp!]
\includegraphics[width=0.5\textwidth]{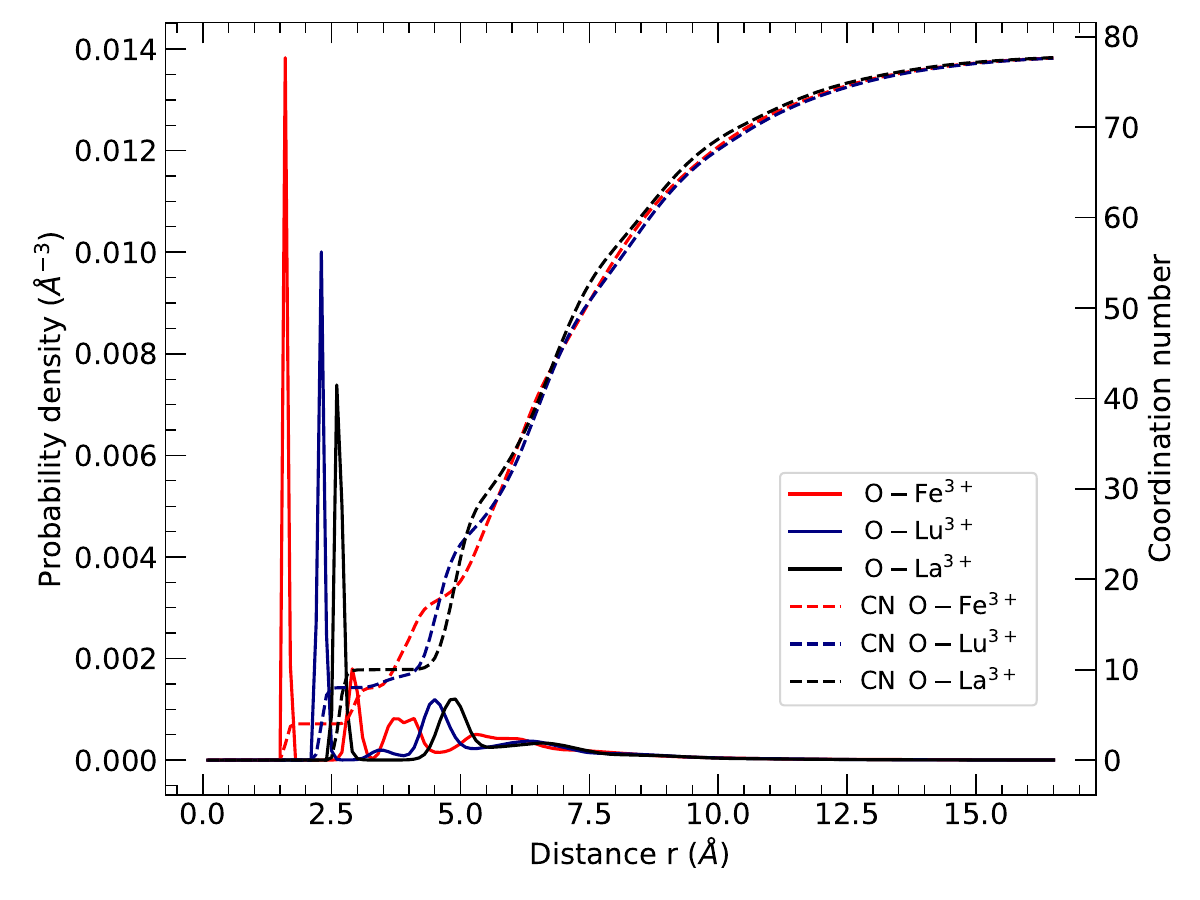} \caption{Radial distribution function (RDF) and coordination number between oxygen sites of \ce{H2O} and \ce{Fe^{3+}} ion (red colored),
\ce{Lu^{3+}}-O (blue colored), and \ce{La^{3+}}-O (black colored) in 
$N_{\ce{H2O}} = 78$ at $T=190$~K.} 
\label{fig:rdf-Oh2-trivalentions}
\end{figure}

The RDFs between the oxygen sites and a trivalent metal ion in $N_{\ce{H2O}} = 78$ are shown 
in Fig.~\ref{fig:rdf-Oh2-trivalentions} at $T =190$~K. 
The location of the maxima and the coordination numbers for \ce{Fe^{3+}}-O, \ce{La^{3+}}-O, and
\ce{Lu^{3+}}-O in $N_{\ce{H2O}} = 78$ are summarized in Table~S4 in SI.
A similar RDF to Fig.~\ref{fig:rdf-Oh2-trivalentions}~(a) but for 
$N_{\ce{H2O}} = 242$  is shown in Fig.~S4 in SI.
As expected, the RDFs indicate more structured hydration shells surrounding \ce{Fe^{3+}}, which is the smallest among the studied ions.
The intensity of the first peak for the \ce{Fe^{3+}}-O is distinctly higher than that of \ce{Lu^{3+}}-O
and \ce{La^{3+}}-O, which indicates the rigidity in the first hydration shell of \ce{Fe^{3+}}
relative to that of \ce{Lu^{3+}} and \ce{La^{3+}}.
The maximum of the \ce{Fe^{3+}}-O peak is at 1.6~{\AA}, while of \ce{Lu^{3+}} and \ce{La^{3+}} at 
2.3~{\AA} at and 2.6~{\AA}, respectively.
It is interesting to note that for \ce{Fe^{3+}} there are two
consecutive peaks between 2.5~{\AA} to 4.5~{\AA}, which indicates that the second and third hydration shells of  \ce{Fe^{3+}} are densely arranged around the \ce{Fe^{3+}} center. Similar structure appears for \ce{Lu^{3+}} but with lower intensity. For \ce{La^{3+}}, which is the largest ion, only one maximum appears at 4.8~{\AA}. 
Figure~S5 in SI for $N_{\ce{H2O}} = 40$ at $T=130$~K, 
shows a rigid structure within the clusters
arising from the immobility of the \ce{H2O} at the low temperature. 

\begin{figure}[htb!]
\centering
\begin{subfigure}[htbp]{0.5\textwidth}
\centering
\includegraphics[width=\textwidth]{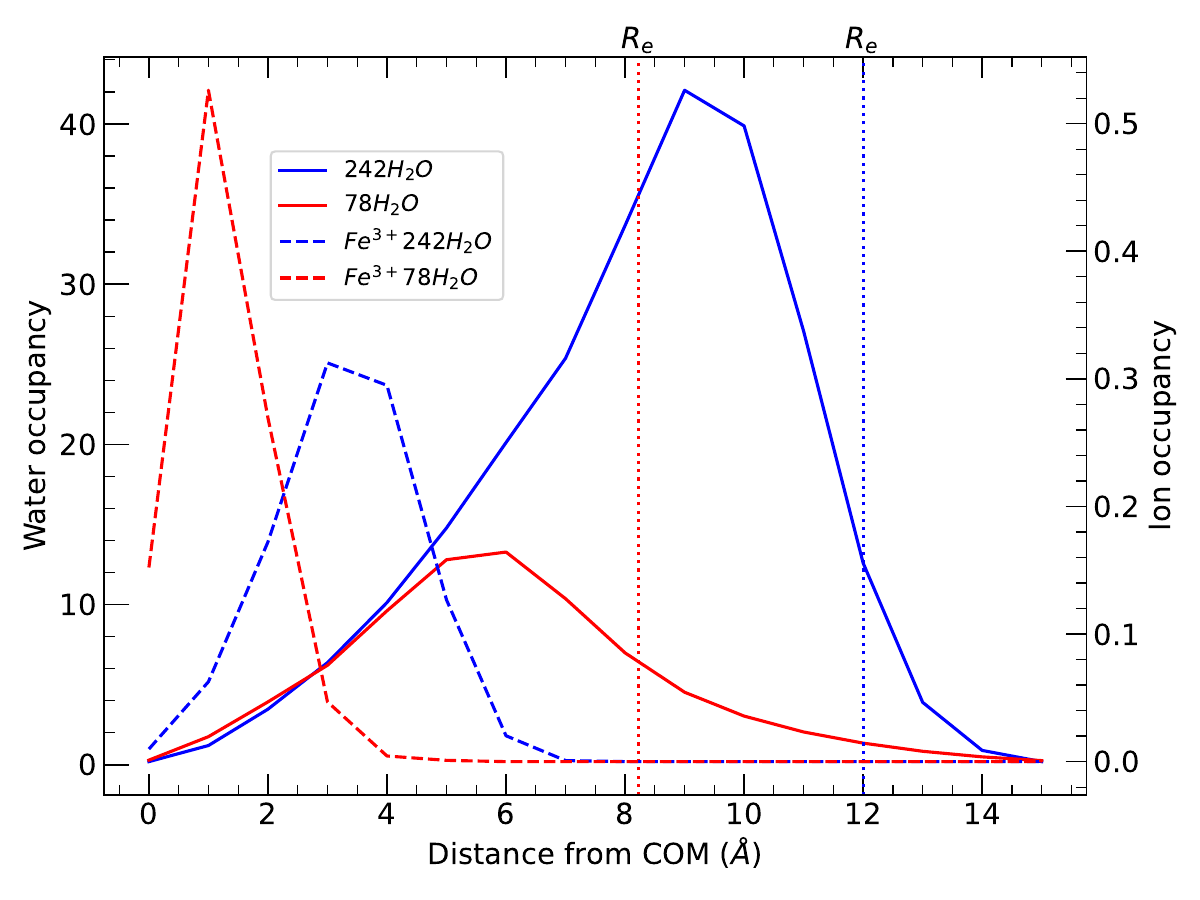}
\caption{}
\end{subfigure}
\begin{subfigure}[htbp]{0.5\textwidth}
\centering
\includegraphics[width=\textwidth]{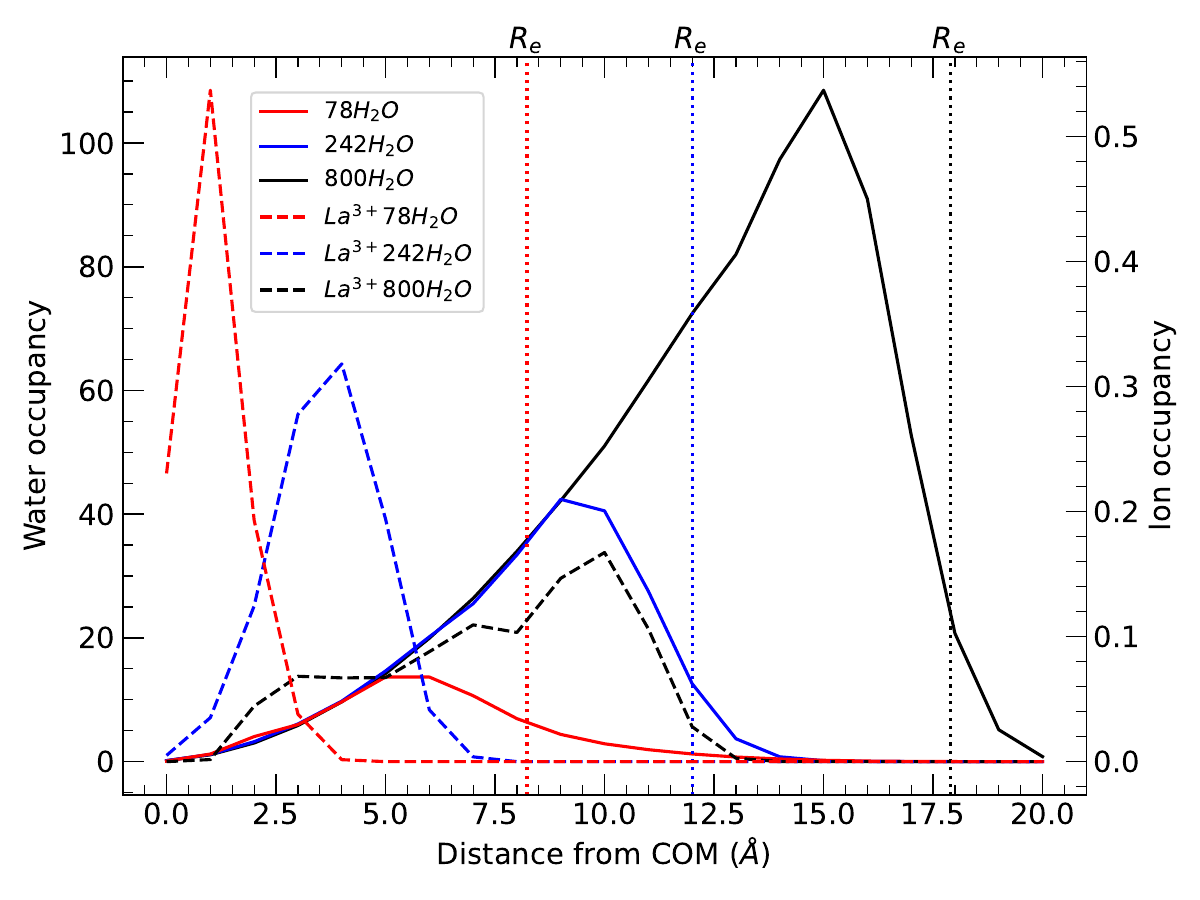}
\caption{}
\end{subfigure}
\caption{Raw data of the probability of encountering the ion and the \ce{H2O} molecules 
at a certain distance from the
cluster's COM. The data have not been divided by the volume
element. The vertical dotted lines mark the equimolar radius ($R_e$) of the
cluster. (a) \ce{Fe^{3+}} in $N_{\ce{H2O}}= 78, 242$, 
(b) \ce{La^{3+}} in $N_{\ce{H2O}}= 78, 242, 800$.  
}
    \label{fig:depth}
\end{figure}

The orientation of the \ce{H2O} molecules is better analyzed when the depth of the ion from the surface
of the cluster is known. Figure~\ref{fig:depth}~(a) shows the raw data of the probability of encountering the ion and the \ce{H2O} molecules
at a certain distance from the
cluster's COM for \ce{Fe^{3+}} in $N_{\ce{H2O}}= 78$ and $N_{\ce{H2O}}= 242$. In $N_{\ce{H2O}}= 78$, \ce{Fe^{3+}}
is found two hydration layers below the cluster surface while in $N_{\ce{H2O}}= 242$ the depth is three 
layers\cite{ng2022iron}
because of the large available volume. The long tail up to 14.0~{\AA} in the probability of the \ce{H2O} 
in $N_{\ce{H2O}}= 78$ 
is a signature of the conical protrusions, while this tail is much shorter in $N_{\ce{H2O}}= 242$. 
Similar to Fig.~\ref{fig:depth}~(a), Figure~\ref{fig:depth}~(b) shows that \ce{La^{3+}} in $N_{\ce{H2O}}= 242$
and $N_{\ce{H2O}}= 800$ is found three \ce{H2O} layers below the surface.

\begin{figure*}[htb!]
\centering
\begin{subfigure}[htbp]{0.3\textwidth}
\centering
\includegraphics[width=\textwidth]{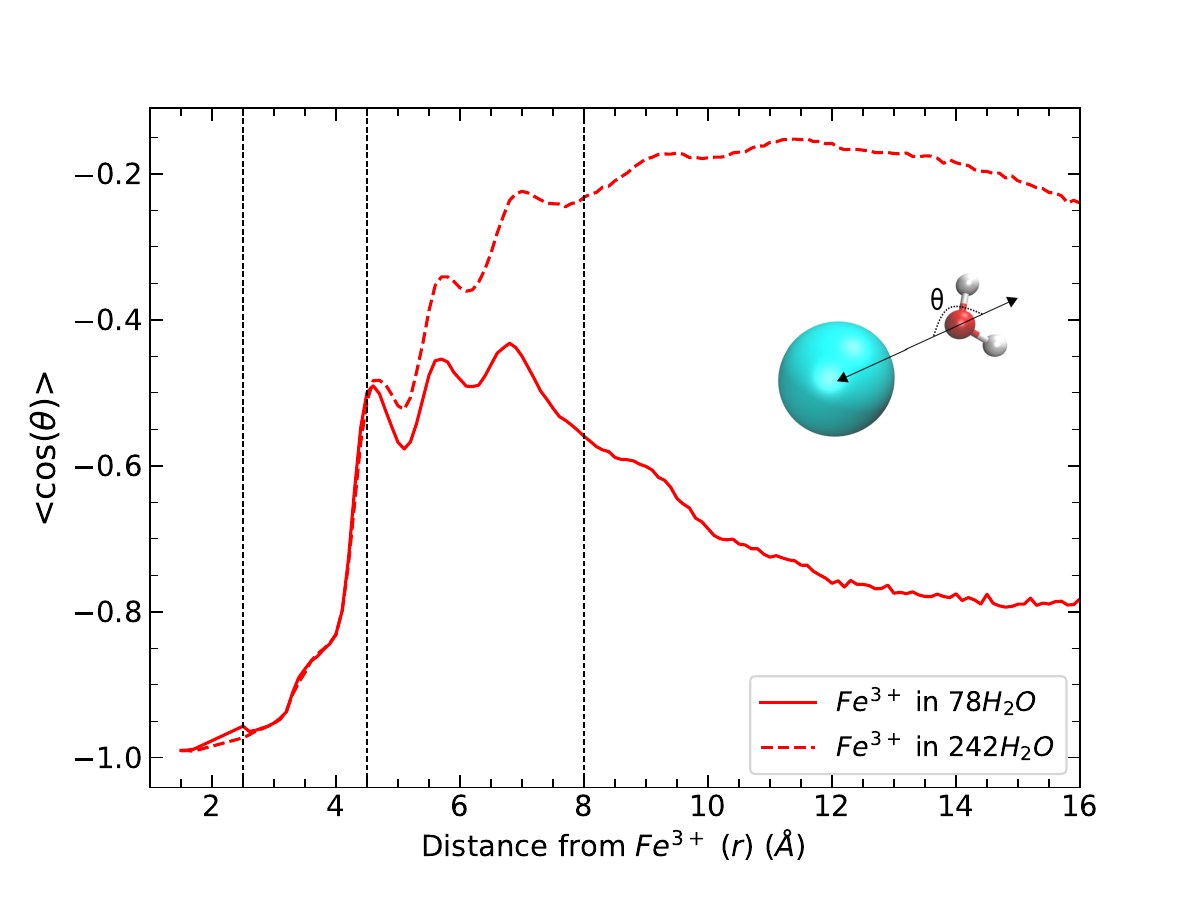}
\caption{}
\end{subfigure}
\begin{subfigure}[htbp]{0.3\textwidth}
\centering
\includegraphics[width=\textwidth]{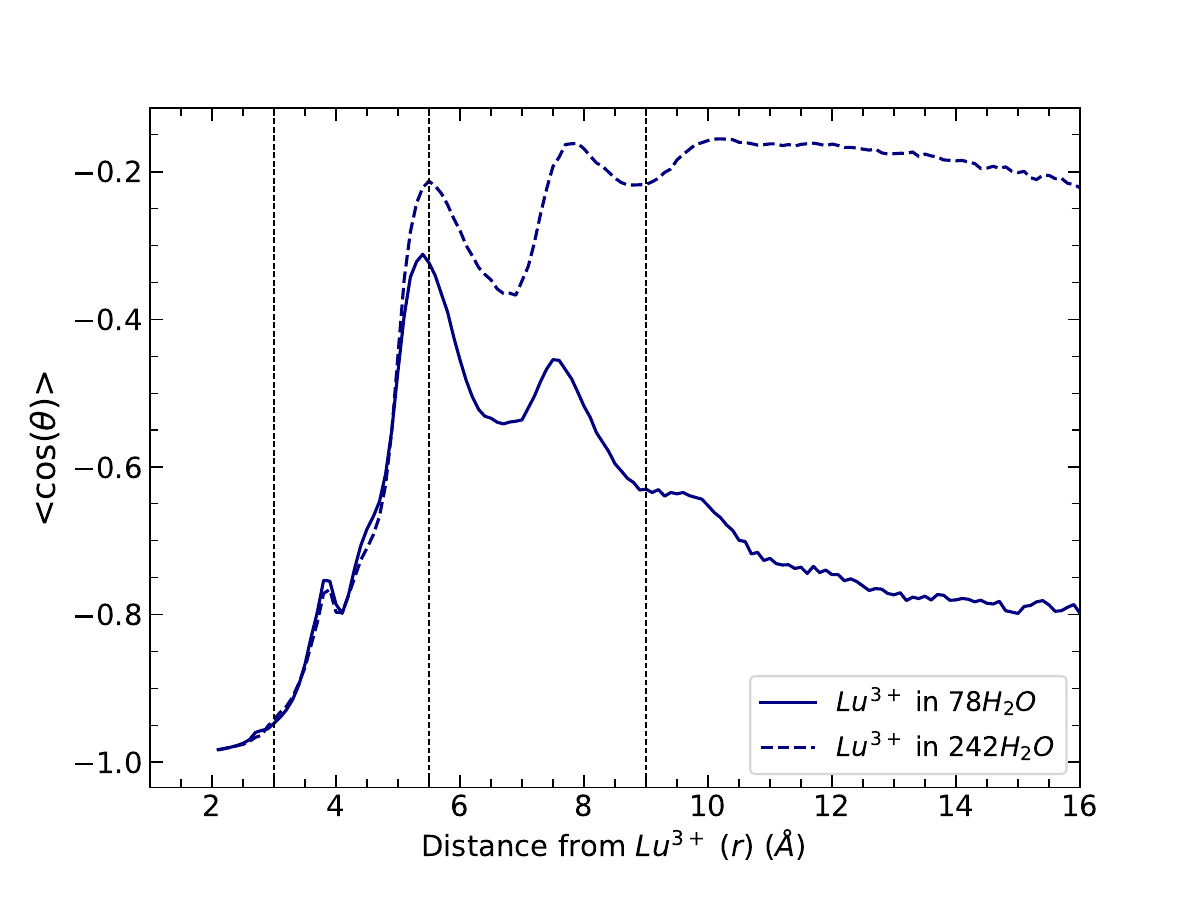}
\caption{}
\end{subfigure}
\begin{subfigure}[htbp]{0.3\textwidth}
\centering
\includegraphics[width=\textwidth]{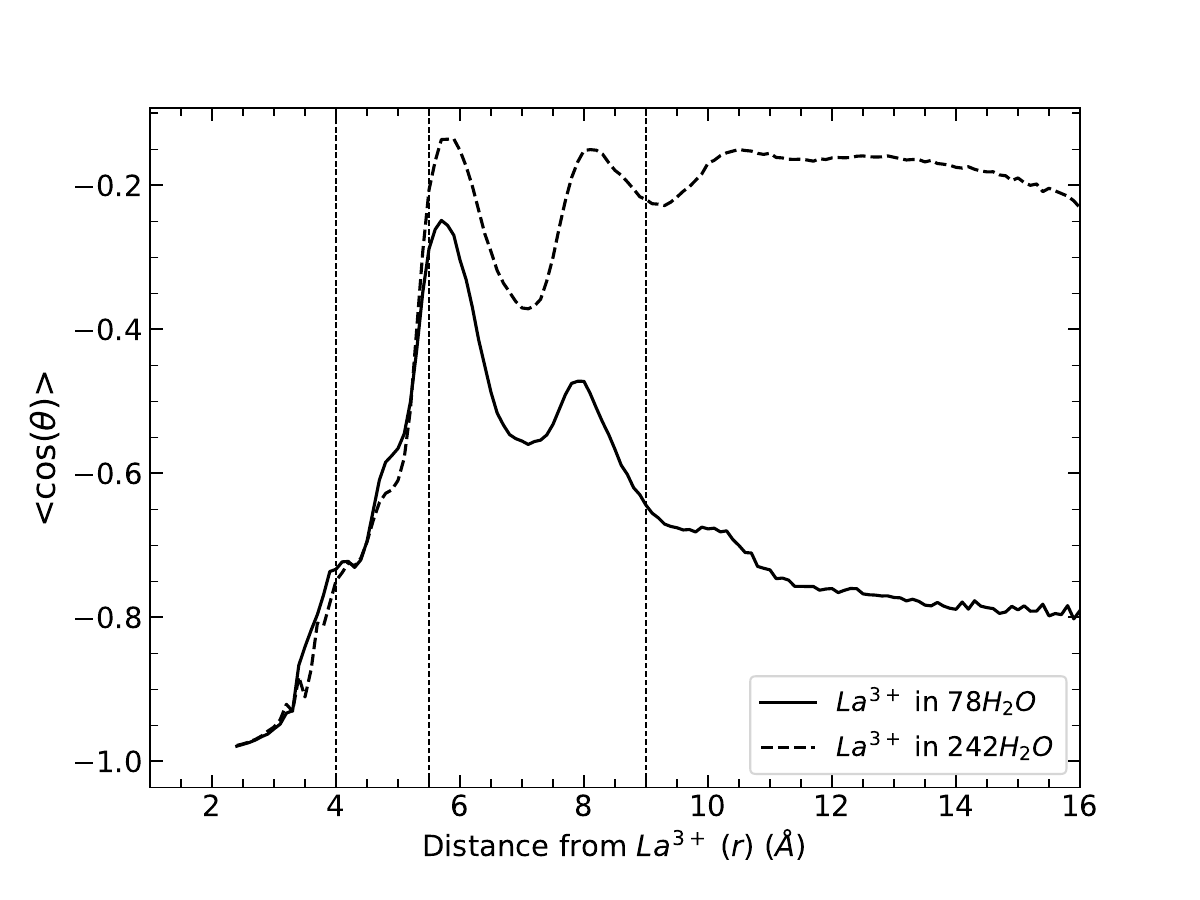}
\caption{}
\end{subfigure}
\begin{subfigure}[htbp]{0.3\textwidth}
\centering
\includegraphics[width=\textwidth]{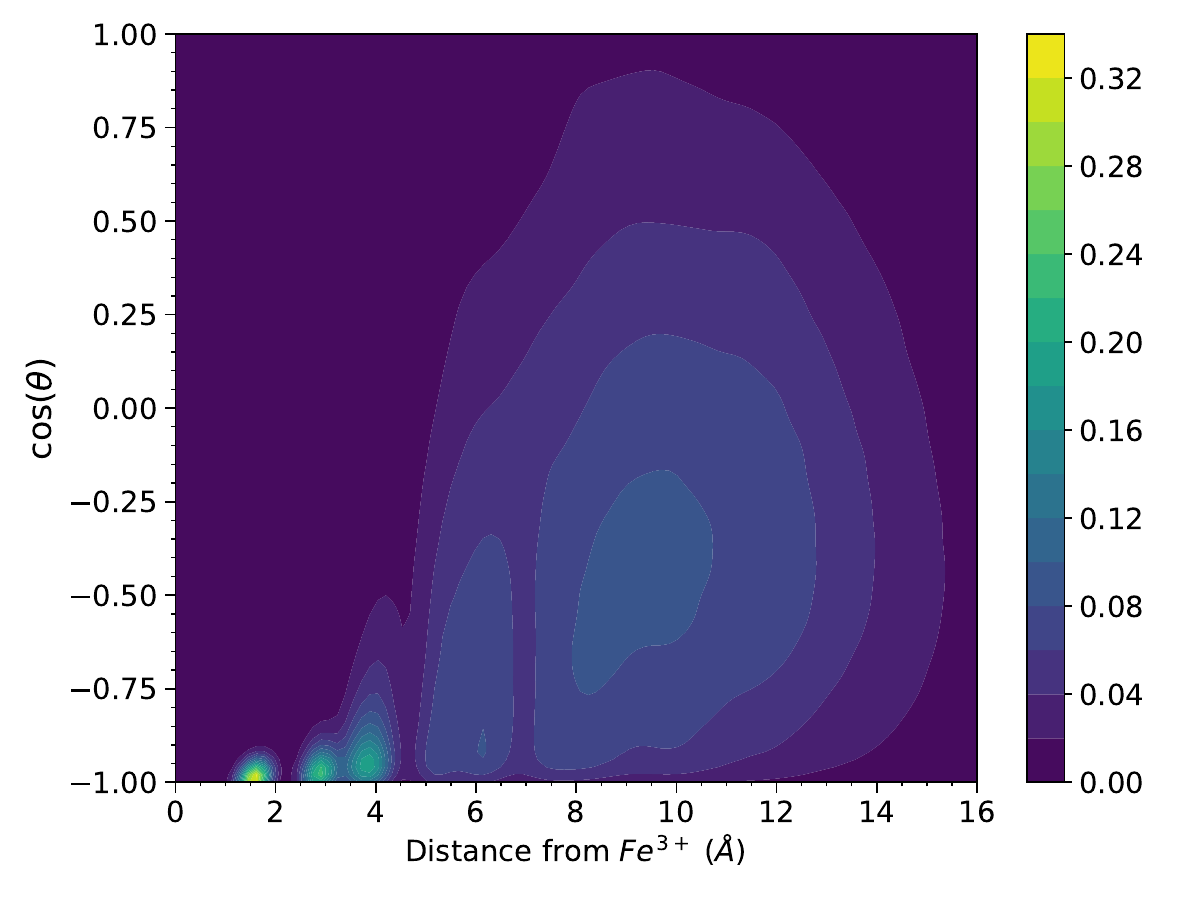}
\caption{}
\end{subfigure}
\begin{subfigure}[htbp]{0.3\textwidth}
\centering
\includegraphics[width=\textwidth]{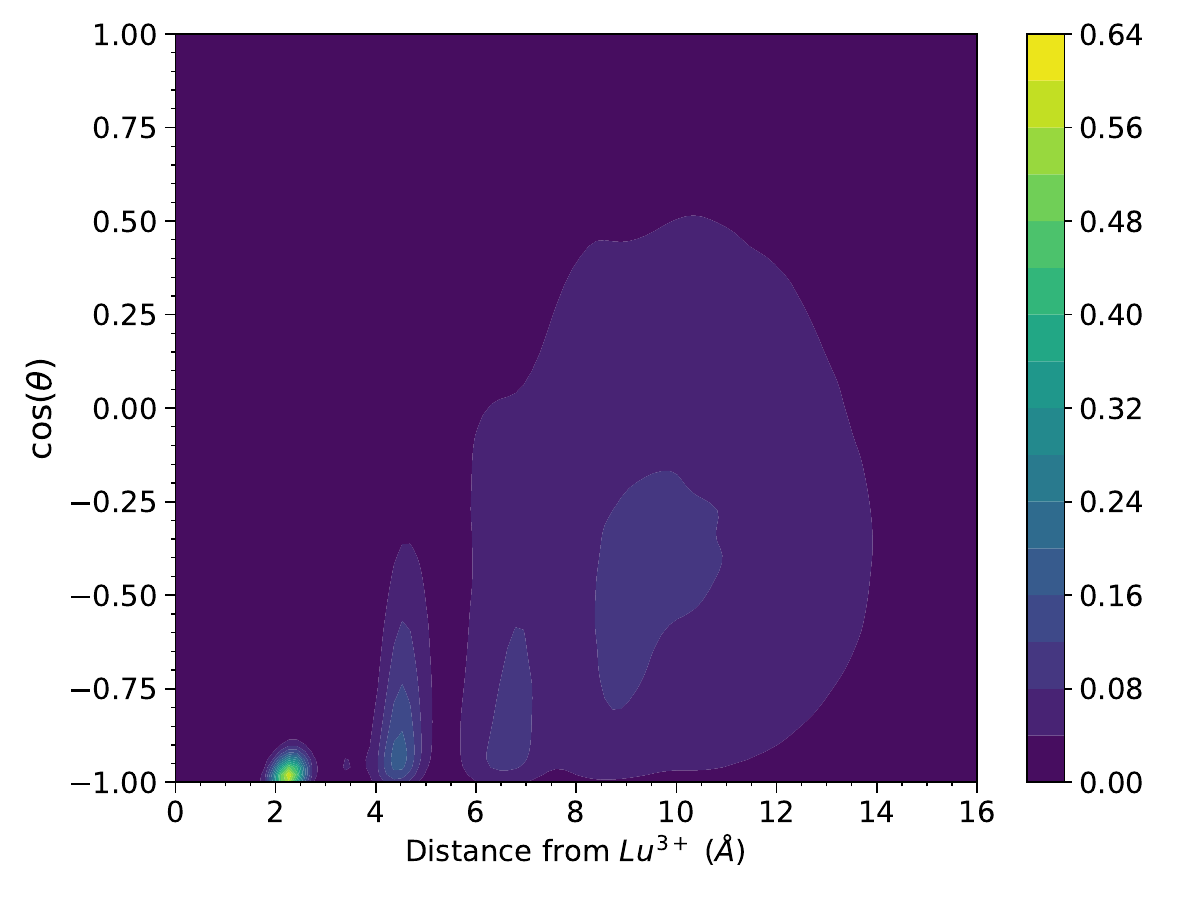}
\caption{}
\end{subfigure}
\begin{subfigure}[htbp]{0.3\textwidth}
\centering
\includegraphics[width=\textwidth]{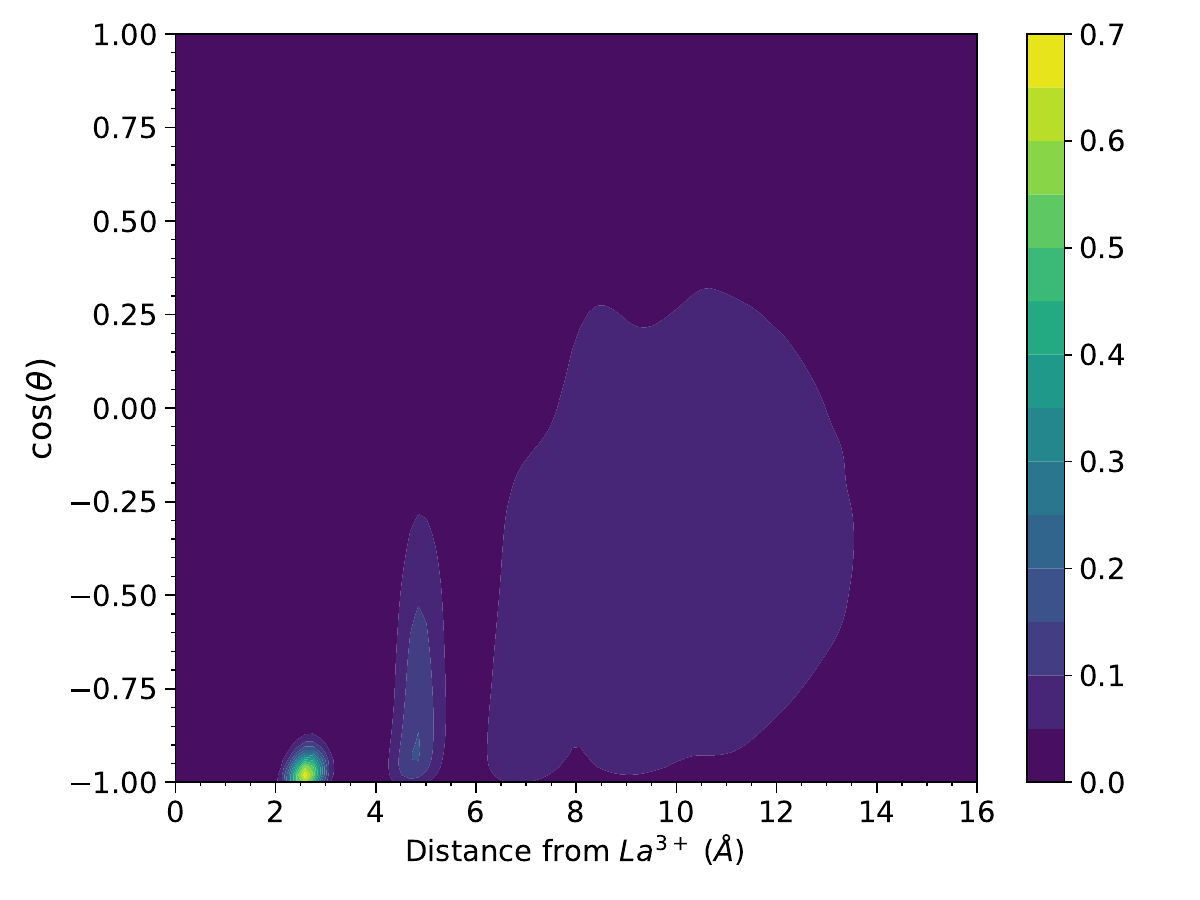}
\caption{}
\end{subfigure}
\caption{(a)-(c) Average $\cos(\theta)$ as a function of distance ($r$) from (a) \ce{Fe^{3+}}, (b) \ce{Lu^{3+}}, 
and (c) \ce{La^{3+}} for $N_{\ce{H2O}} = 78$ (solid lines) and $N_{\ce{H2O}} = 242$ (dashed lines) at 190~K. $\theta$ is defined as the angle between the dipole moment of a \ce{H2O} molecule defined from the negative site to the positive site and  a unit vector from the oxygen site of a \ce{H2O} molecule to the center of the ion
(see inset in (a)). The vertical lines mark the boundaries of the hydration shells as estimated
from the RDFs in Fig.~\ref{fig:rdf-Oh2-trivalentions}. 
(d)-(e) Contour maps of $\cos(\theta)$ distribution of \ce{H2O} dipole orientations
within successive spherical shells centered 
at (d) \ce{Fe^{3+}}, (e) \ce{Lu^{3+}}, and (f) \ce{La^{3+}}
in $N_{\ce{H2O}} = 242$ at 190~K. Contour maps for $N_{\ce{H2O}} = 78$ are shown in Fig.~S6 in SI.}
\label{fig: avgCos_theta_Nh2o-190K}
\end{figure*}

\begin{figure}
    \centering
    \includegraphics[width=1.0\linewidth]{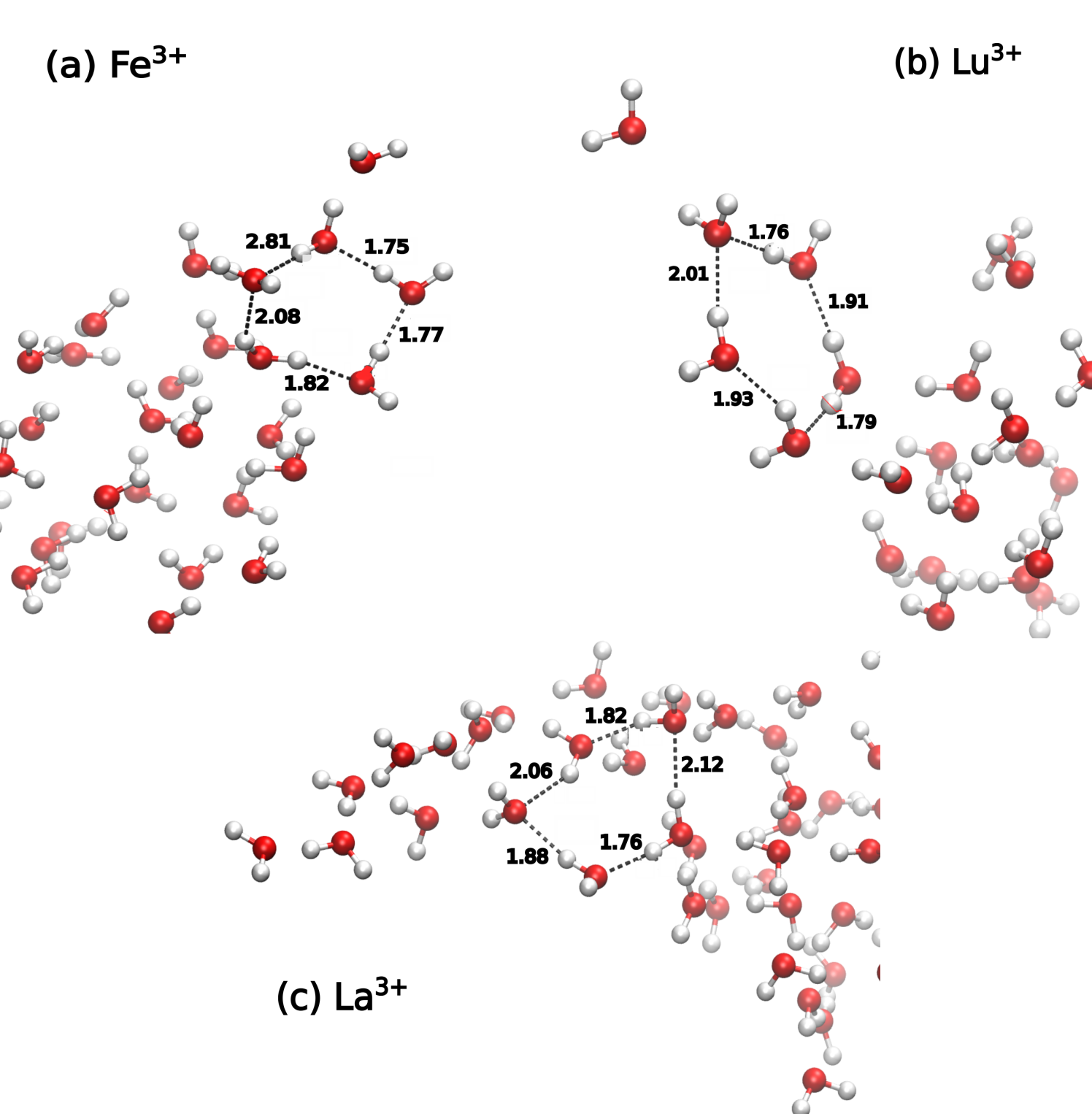}
    \caption{Typical snapshots of the water ring structure in one of the conical
      protrusions of (a) \ce{Fe^{3+}}, (b) \ce{Lu^{3+}}, and (c) \ce{La^{3+}} in $N_{\ce{H2O}}=78$ at $T =190$~K. In each snapshot, the lengths of H-bonds are indicated.}
    \label{fig:ring-ion-78h2o}
\end{figure}

Figure~\ref{fig: avgCos_theta_Nh2o-190K}~(a), (b), and (c) show the orientation of the \ce{H2O} molecules 
around \ce{Fe^{3+}}, \ce{Lu^{3+}}, and \ce{La^{3+}}, respectively,
for $N_{\ce{H2O}}= 78$ and $N_{\ce{H2O}}= 242$. 
In the first hydration shell of the ion, there is not only a strong local ordering as indicated by the
RDFs (Fig.~\ref{fig:rdf-Oh2-trivalentions}) but also strong orientational
ordering 
of the \ce{H2O} dipoles where $\theta \sim 180^{\circ}$ (definition of $\theta$ is found in
Fig.~\ref{fig: avgCos_theta_Nh2o-190K}~(a)) as shown in the first 
segment of Fig.~\ref{fig: avgCos_theta_Nh2o-190K}~(a), (b), (c).

In the second hydration shell, the orientation becomes rapidly more disordered. As it is expected,
the larger the trivalent ion is, the more the randomization towards a zero value of 
$\langle \cos(\theta) \rangle$ is. In the values of $\cos(\theta)$ 
within the second hydration shell of the \ce{Fe^{3+}} and \ce{Lu^{3+}} we see a distinct peak that
corresponds to the two peaks that we see in their RDFs (Fig.~\ref{fig:rdf-Oh2-trivalentions})  
between 2.5-4.5~{\AA} for \ce{Fe^{3+}} and 3.0-5.5~{\AA} for \ce{Lu^{3+}}. 
  The pattern of \ce{H2O} dipole moment orientations follows: the stronger electric field 
  exerted from the ion on the \ce{H2O} molecules, the more oscillations 
  of $\langle cos(\theta) \rangle$ occur.
  This effect is more pronounced for \ce{Fe^{3+}} than for \ce{Lu^{3+}} or \ce{La^{3+}}. 
The ion's strong electric field even though decreases in the third hydration shell still persists enough to induce a slight water orientation. 
  In clusters with up to several tens of \ce{H2O} molecules, as for
  example in $N_{\ce{H2O}}= 78$, the third hydration shell surrounding the ion
is frequently not continuous because of the ``star''-shaped 
configurations that form troughs near the ion (see Fig.~\ref{fig:78tip3-Fe3+-190K}). 
For \ce{La^{3+}} and \ce{Lu^{3+}} the orientation of the \ce{H2O} molecules
in the third and later hydration shells becomes more organized, while for \ce{Fe^{3+}}
the orientation is obviously more aligned beyond the third hydration shell.
The increased orientation beyond the second hydration shell
is attributed to the alignment of the \ce{H2O} 
molecules within the conical rays of the star-like shapes. Typical snapshots that show the alignment of the \ce{H2O} molecules is shown in Fig~\ref{fig:ring-ion-78h2o}. In Fig~\ref{fig:ring-ion-78h2o}, frequently encountered ring structures formed by the \ce{H2O} molecules are indicated. Such 
structures were not as likely to appear in $N_{\ce{H2O}}= 242$ clusters. As it will 
be discussed in the next section, 
in the proxy model system containing three \ce{H3O+} ions instead of a 
trivalent metal ion, 
these ring structures show increased proton diffusion.
In $N_{\ce{H2O}}= 242$ clusters, the degree of disorder in the third hydration shell 
and beyond is much higher than in $N_{\ce{H2O}}= 78$ because the instability 
has ceased in this cluster size.
The orientation of the \ce{H2O} molecules is not influenced anymore by the
formation of the conical shapes but it may be influenced by distinct shape fluctuations as 
indicated 
in Fig.~\ref{fig:moI-fe3+-190K-70h2o}. 
An average $cos(\theta)$ near zero indicates no preferential alignment of water dipoles, 
reflecting a bulk-like water orientation.

A detailed presentation of the distribution of the $\cos(\theta)$ as a function of the distance from
the ion is shown in Fig.~\ref{fig: avgCos_theta_Nh2o-190K}~(d), (e), and (f) for $N_{\ce{H2O}}= 242$,
while for $N_{\ce{H2O}}= 78$ is shown in Fig.~S6 in SI. In Fig.~\ref{fig: avgCos_theta_Nh2o-190K}~(d)-(e)
distinct maxima in the range 0-6.0~{\AA} are present for the three ions. 
In the range of 7.5-10.5~{\AA} for \ce{Fe^{3+}}, and in 8.0-10.5~{\AA} for \ce{Lu^{3+}} a broad
local maximum appears that is not present for \ce{La^{3+}}. Any of the trivalent metal ions
studied here, resides the
most of the time approximately three \ce{H2O} layers below the cluster surface as shown in
Fig.~\ref{fig:depth}. 
As a result these broad local maxima also include the orientation of surface molecules.
The presence of these local maxima are signatures of the
the range of the ion's electric field. They indicate that for \ce{Fe^{3+}} and
\ce{Lu^{3+}} the range of the electric field extends to three hydration shells
around the ion.

An analysis of the distribution of the acceptor (A)-donor (D) relation of \ce{H2O} 
molecules using VMD in $N_{\ce{H2O}}=78$ at $T =190$~K shows that relative to pure
\ce{H2O} clusters, the percentage of ADD and AADD in the metal ion clusters 
decreases with decreasing ion size, D, DD, consistently 
increases with decreasing ion size, while A increases by the same amount for
all ion sizes. In the analysis the definition of the H-bond was based on 
O-O distance equal to 3.2~{\AA} and angle $30^\circ$.
The decrease of ADD and AADD with decreasing ion size is expected because
of the larger disruption of the H-bonded network by the smallest ion, while the increase of 
D, DD, and A, may reflect the ordering of the \ce{H2O} molecules in the
conical protrusions. 

Based on these findings, we expect the spectrum of \ce{Fe^{3+}}-\ce{H2O} clusters 
to exhibit a significant red shift in the free-OH band compared
to the monovalent and divalent ions studied by Williams
et al.\cite{prell2011structural, bush2008reactivity}, due to the 
intense electric field of \ce{Fe^{3+}}. The morphology observed in our simulations,
with highly anisotropic structures for small clusters and a transition toward 
more spherical configurations beyond $\sim 100$ \ce{H2O} molecules, 
suggests that the intensity and position of the IR bands could vary with cluster size. 
Specifically, smaller clusters should exhibit a broader and more structured spectrum 
due to variable solvation environments due to conical protrusions, 
whereas larger clusters may display a more defined free-OH band shifted to lower energies, 
reflecting enhanced hydrogen-bond stabilization. 
Additionally, the bonded-OH band is expected to extend to lower frequencies 
compared to systems with lower-charge ions, indicating a strong reorganization of the 
hydrogen-bond network in response to \ce{Fe^{3+}} electric field.
A similar trend is expected for \ce{Lu^{3+}} and \ce{La^{3+}}
though less pronounced due to their larger ionic radius.

\subsection{Acidity within protruding rings in multi-protonated aqueous clusters}
\begin{figure}[h] 
\centering    
\includegraphics[width=0.47\textwidth]{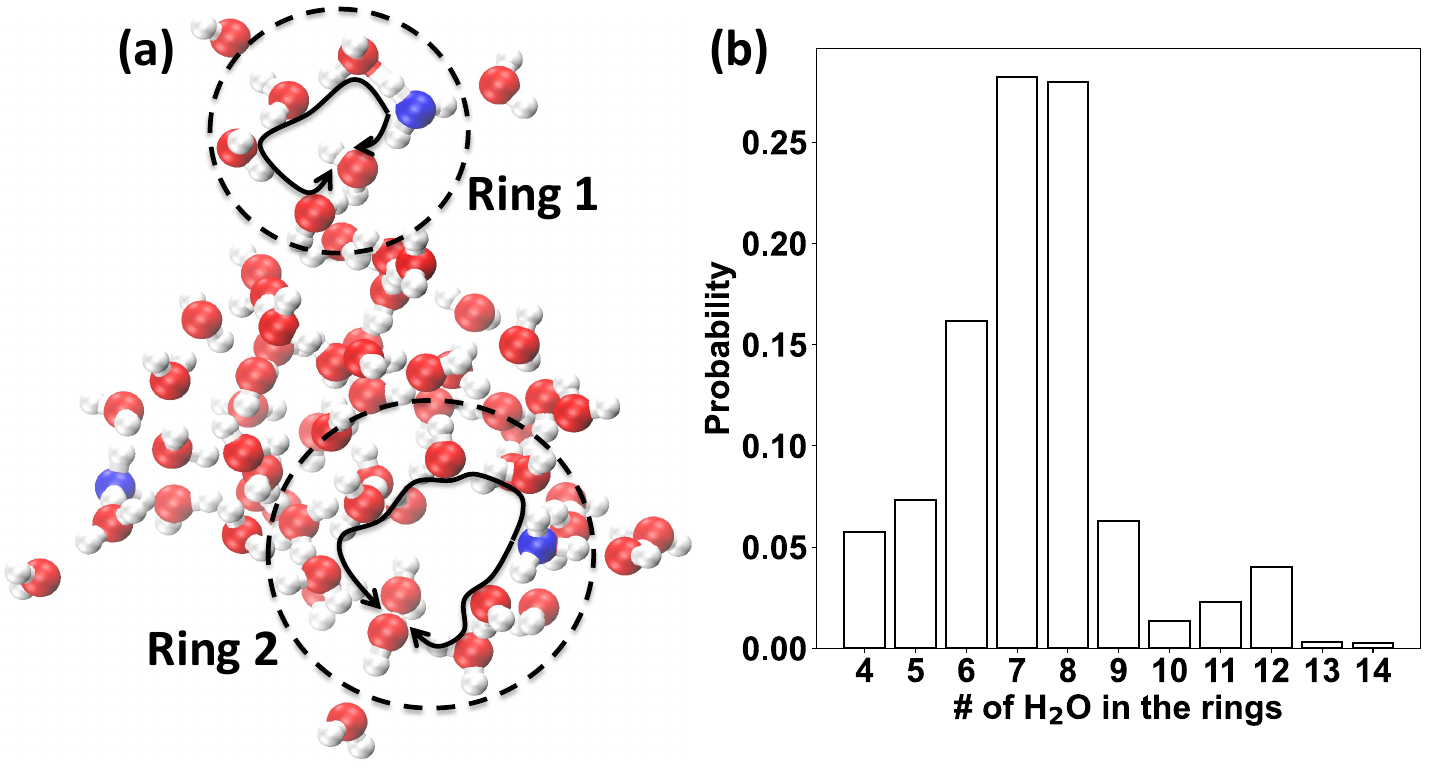}
\caption{(a) A typical snapshot from the AIMD trajectory of the 64~\ce{H2O}~-~3~\ce{H3O+} cluster. Oxygen sites of the \ce{H2O} molecules are colored red and H sites are white. The O sites of the three $\mathrm{H}_3 \mathrm{O}^{+}$ ions are colored blue. Here, the example of two rings (out of three) of different length that contain a single \ce{H3O+} ion in each of them are shown within the dashed circles. The $\mathrm{H}_3 \mathrm{O}^{+}$ ions are characterized by outgoing chains, as indicated by the path marked by the black arrow. The length of ring 1 is 6 containing a single \ce{H3O+} ion and 5 other \ce{H2O} molecules connected with one another by hydrogen bonds. Ring 2 is composed of 8 water molecules, containing a single \ce{H3O+} ion and 7 other \ce{H2O} molecules connected with one another by hydrogen bonds. (b) Distribution of ring sizes around the three $\mathrm{H}_3 \mathrm{O}^{+}$ ions.}
\label{c64-3h3o-ring}
\end{figure}

To probe the acidity in the conical protrusions formed by the instability, we employ multi-protonated aqueous clusters as a model system. AIMD is employed which allows for direct observation of proton transfer events. Because of computational efficiency a cluster comprised 
$N_{\ce{H2O}} = 64$ and 3\ce{H3O+} ions is modeled. This cluster is slightly above the Rayleigh limit ($X > 1$ in Eq.~1) and comparable in size with the cluster sizes manifesting instability in the presence of a trivalent ion. A typical snapshot of a $N_{\ce{H2O}} = 64$-3\ce{H3O+} cluster is shown in Fig.~\ref{c64-3h3o-ring}~(a). The three \ce{H3O+} are located on the surface as it has been found for single \ce{H3O+} ions in clusters\cite{petersen2004hydrated,zhang2024double}. In this cluster an extrusion has been formed on the surface that contains a single \ce{H3O+} ion.
Because of supercharging the cluster is expected to 
divide easily into two or three sub-clusters, where each of them carries a solvated \ce{H3O+} ion. 
Indeed in classical MD we observed fragmentation into three sub-clusters within a few ps. 
AIMD simulations showed the $N_{\ce{H2O}} = 64$-3\ce{H3O+} cluster to 
be connected with an extruded segment from the body of the cluster.
No division of the cluster took place within 7~ps.

The difference in the fragmentation
time 
observed between MD with empirical force fields
and AIMD may arise 
from their distinct capabilities in describing intermolecular 
interactions and the associated cluster dynamics. In ``classical'' 
MD simulations, the \ce{H3O+} ions are typically treated using a rigid water 
model inherently neglects the effects of electronic polarization and charge redistribution, 
leading to a weaker and less stable hydrogen bonding network. 
Consequently, the \ce{H3O+} ions in classical MD are more prone to cluster fragmentation. 
By construction, AIMD incorporates the explicit calculation of electronic structure, allowing for dynamic charge transfer, polarization, and hydrogen bond reorganization within the cluster. These features significantly enhance the structural stability and cohesion of the clusters. Furthermore, the weakness
of empirical force fields to inherently capture proton transfer between water molecules 
contrasts with AIMD, where such processes naturally emerge, 
contributing to the overall stabilization of the clusters.

Molecular modeling of droplets containing a few tens of thousands of
\ce{H2O} molecules has shown that 
near the onset of instability, clusters containing several single ions form conical protrusions on the surface\cite{kwan2022conical}. The cones formation is governed by the same physical principles as those in the presence of a trivalent ion. 
The conical protrusions are dynamic, they can form and retract void of any ions, 
or when a conical deformations forms, an ion at the vicinity of the cone may enter its base 
and diffuse to the tip from where it is emitted\cite{consta2022atomistic}. 
In the $N_{\ce{H2O}} = 64$-3\ce{H3O+} cluster the extrusion is maintained within 7~ps of AIMD run. Because of the small size of the cluster modeled by AIMD the extruded segment contains only several \ce{H2O} molecules that are highly oriented. Strong orientation of the \ce{H2O} molecules has been also found in the conical protrusions
in the metal ion clusters at the instability 
regime as was discussed in 
Fig.~\ref{fig: avgCos_theta_Nh2o-190K}  and Fig.~\ref{fig:ring-ion-78h2o}). 

\begin{figure}[h]
\centering    
\includegraphics[width=0.48\textwidth]{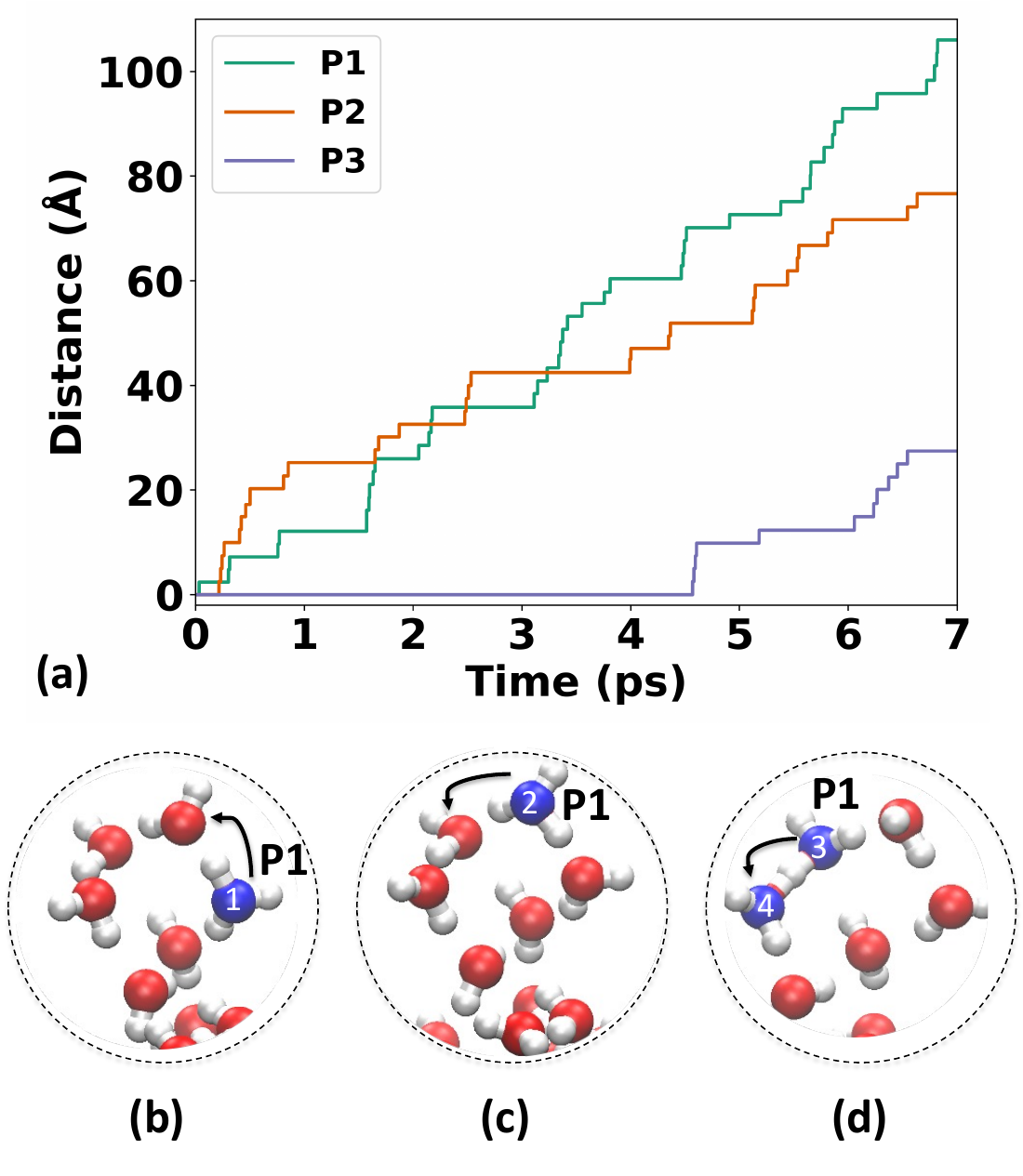}
\caption{(a) The accumulative distance of proton transfer for three protons (P1-P3) in the 64~\ce{H2O}-3~\ce{H3O+} system. (b)-(d) Typical snapshots of proton (P1) transfer along the rings were taken from the AIMD simulations. The color coding of the atomic sites is the same as that in Fig.~\ref{c64-3h3o-ring}. The black dashed circles in (b)-(d) highlight the configuration facilitating the fast proton transfer from oxygen sites of 1 to 4 (colored in blue) within the ring.}  
\label{c64-3h3o-distance}
\end{figure}
In particular, Fig.~\ref{fig:ring-ion-78h2o} shows the formation of ring H-bonded structures observed in
the conical regions for the three ions studied here. 
Similarly, Fig.~\ref{c64-3h3o-ring}~(b) indicates that in the extruded segment, and around one of the \ce{H3O+} ions in the more spherical body of the cluster, \ce{H2O} molecules form H-bonded rings. 
The ring structure was determined based on setting up the cutoffs between the O-O distance to 3.2 \AA, then the H-O distance to 2.39 \AA, and then the hydrogen bonding angle O-H-O to $\geq$ $155^\circ$. The 
program runs on the searching rule of all potential proton transfer pathways starting 
from \ce{H3O+} ion, then it progressively finds the closed ring step by step using the 
cutoffs as defined above. 
To make the ring structure search more efficient, a preliminary search cutoff 
of 3.5 \AA\ is added. In the way, the present water molecule 
will search the next candidate water molecule that is likely to participate 
in ring structure within this preliminary region.

It is noted that all ring structures such as those 
shown in Fig.~\ref{c64-3h3o-ring}~(a) of rings 1 and 2 can be viewed as rings 
departing from the \ce{H3O+} ion, with the black arrows in the 
figure pointing out the possible pathways for proton transfer. 
It was also found that the size of the rings around \ce{H3O+} changed during the simulation. 
Theoretically, a \ce{H3O+} ion can exist with three ring structures\cite{wu2008improved,hassanali2013proton}, each starting with one of the water molecules and ending with one of the neighboring water molecules, forming a closed ring. The rings that more frequently
appear in the $N_{\ce{H2O}} = 64$-3\ce{H3O+} cluster 
over the 7~ps AIMD run are mainly composed 
of 5 to 8 \ce{H2O} molecules. This is shown in Fig.~\ref{c64-3h3o-ring}~(b). 
Snapshots of longer ring structures with 10 and 12 \ce{H2O} molecules
are shown in Fig.~S7 in SI.  Ring structures of \ce{H2O}
have been identified
in previous works by Hassanali et al.~\cite{hassanali2013proton} within the
bulk solution. Differently in our study, these rings are found at the surface of the
cluster or in the protruding portion. Hassanali et al.~\cite{hassanali2013proton} have
argued that the units forming the closed rings are related to the
stability of the water network around the protons.
The rings that are made up of a smaller number
of \ce{H2O} molecules (less than 10 found in the work of Ref.~\cite{hassanali2013proton}) 
provide longer trapping time for the proton, while regions of larger rings provide
shorter trapping time because more transfer paths become 
available to the proton due to instability of the larger rings.
As it will be discussed the ring structures, mainly pentamers to heptamers, in the cluster
promote the proton transfer between two adjacent \ce{H2O} molecules and 
proton diffusion over several \ce{H2O} molecules.

To the knowledge of the authors, the acidity in the conical region have not been studied before.  
Clusters comprised $N_{\ce{H2O}} = 64$-3\ce{H3O+} show that during  $\sim 7$~ps in their connected form, proton transfer events occur between the \ce{H3O+} ion and the 
neighboring \ce{H2O} molecules and proton diffusion 
over several \ce{H2O} molecules. The protons in the  $N_{\ce{H2O}} = 64$-3\ce{H3O+} 
cluster are labeled as P1, P2, and P3. Fig.~\ref{c64-3h3o-distance} shows the 
cumulative distance that each proton has traveled within 7~ps, including returns to the 
previous donor \ce{H2O} molecule. Distinct dynamics among the protons found in different 
regions of the cluster is observed. 
P1, that is found in the protrusion undergoes fast 
forward and back transfer between two \ce{H2O} molecules but also successful 
transfers that allow it to diffuse 5 \ce{H2O} molecules away from its 
starting \ce{H2O} molecule. 
Snapshots that focus on the region of the cluster where P1 is found 
is shown in Fig.~\ref{c64-3h3o-distance}~(b)-(d) and a Movie in SI shows the proton diffusion.  
P2 also transfers  over several 4-5 \ce{H2O} molecules away from its parent \ce{H2O} molecule, 
then goes back to the initial starting point after 20~30 fs. 
P1 and P2 participate in ring structures as shown in Fig.\ref{c64-3h3o-ring}~(a)-(b). P3 that is in the more spherical body of the cluster (but still stays on the surface as a pyramid 
structure with three hydrogens facing toward to the cluster's interior) 
shows delocalization between only two or three \ce{H2O} molecules with many 
re-crossings and returns to its parent \ce{H2O} molecule. 
Simulations of ``star-like'' droplets
has shown that a death-birth process of the rays may take 
place\cite{consta2010manifestation}. By analogy, one
may expect that diffusion of P3 toward the \ce{H2O} molecules at the 
bottle-neck may lead to the retraction of the pentameric ring on
the cluster surface, while the ring that contains P2 or P1 may start to protrude.

The proton transfer in the clusters studied here 
has many of the features that have been already identified 
in the bulk solution\cite{tuckerman1994ab,tuckerman1995ab,lobaugh1996quantum,marx2006proton,calio2021resolving,eigen1964proton,zundel1969hydration,berkelbach2009concerted,decoursey2014philosophy,natzle1985recombination,huckel19283,de1806decomposition,cukierman2006tu,agmon1995grotthuss,agmon2016protons,geissler2001autoionization,brewer2001formation,hassanali2011recombination,hassanali2013proton,newton1971ab,markovitch2008special} and liquid \ce{H2O} planar interface in contact with the vapor \cite{vacha2007autoionization,vacha2011orientation,winter2009behavior,ottosson2011increased,mishra2012bronsted,hub2014thermodynamics,wick2009investigating,baer2014toward,mundy2009hydroxide,iuchi2009hydrated,petersen2004hydrated,kumar2013exploring,zhang2024double}. Our simulations show significant delocalization of \ce{H3O+} ions among neighboring water molecules. Specially, this observation is in agreement with earlier multi-state empirical valence bond simulation studies by Voth and coworkers in aqueous clusters\cite{iuchi2009hydrated,petersen2004hydrated,kumar2013exploring}, who demonstrated proton transfer delocalization and frequent transitions between Zundel and Eigen complexes. P3 shows frequent proton delocalization over two or three \ce{H2O} molecules on the surface corresponding to \ce{H5O2+} (Zundel) and the \ce{H7O3+} (chain) complexes. 
In addition, the interconversion between the \ce{H9O4+} (Eigen) and \ce{H5O2+} (Zundel) complexes near the surface of the clusters were observed because a central \ce{H3O+} in the Eigen complex randomly approaches one of the three neighboring \ce{H2O} molecules during simulations. In this way, the proton transfer between one \ce{H3O+} and one \ce{H2O} on the surface of the clusters follows the Eigen-Zundel-Eigen mobility mechanism\cite{markovitch2008special} and was found by \citeauthor{markovitch2008special} on the excess proton in the bulk solution using Car-Parrinello MD simulations.

In summary, it was demonstrated that ring structures in protonated clusters assist proton
transfer and diffusion. Formation of rings was frequently found in the conical protrusions
around \ce{Fe^{3+}}, \ce{Lu^{3+}}, or \ce{La^{3+}} when a cluster is at the instability
regime. The facile proton transfer through the rings suggests higher acidity in the
conical regions. 
Moreover,
the frequent proton delocalization within $\sim 7$~ps of AIMD simulation
suggests that a more realistic coarse-grained model of \ce{H3O+} ion in 
modeling \ce{H3O+}-macromolecule interactions in droplets may include the 
delocalization of the proton charge over 2 or 3 \ce{H2O} molecules.

\section{Conclusion}

We addressed the following key questions:
(a) How does charge-induced instability manifest in sub-nanometer-sized aqueous clusters charged with a single trivalent metal ion?
(b) How do the electric field of the metal ion and the cluster geometry (e.g., conical features) influence the orientation of surrounding \ce{H2O} molecules?
(c) What is the acidity of water molecules located in the conical protrusions on the surface of the cluster?

The first question explores how a fundamental physical phenomenon--charge-induced instability--
transitions from minute nanoscopic systems, where macroscopic theories typically fail, 
to microscopic behavior. Our simulations revealed that aqueous clusters charged with a trivalent ion in the instability regime acquire morphologies with multiple conical protrusions on their surfaces. The number of protrusions decreases progressively with increasing cluster size.
Unlike mesoscopic and microscopic droplets--where `star-like' shapes transition 
sharply with droplet radius--the shape transitions in sub-nanometer clusters occur continuously. 
The cluster evolves through triangular, elongated two-point, single-point, and 
more spherical configurations, often exhibiting surface cones along the way.
We propose that the continuous nature of these shape transitions 
stems from the large relative shape fluctuations in sub-nanometer clusters, 
which reduce the energy barriers between different configurations. 
The ion size also influences cluster morphology: for example, 
for a fixed number of \ce{H2O} molecules, \ce{Fe^{3+}}-containing clusters more frequently adopt elongated two-point configurations, whereas \ce{La^{3+}}-containing clusters favor single-point structures.

We found that the electric field of the trivalent ion, regardless of its size, 
strongly influences its immediate two hydration shells. 
Furthermore, for clusters undergoing instability, the collective electric field 
from the ion and surrounding \ce{H2O} molecules beyond the second hydration 
shell strongly orients the \ce{H2O} molecules, in contrast to those in clusters that 
are not subject to instability. 
In clusters where instability effects have subsided, the trivalent ion is 
located approximately three hydration shells below the surface. 
The orientation of the third hydration shell--which includes surface \ce{H2O} molecules--is 
differently influenced by \ce{Fe^{3+}}, \ce{Lu^{3+}}, and \ce{La^{3+}}, depending on the ion's size.

The cluster morphologies observed in our simulations—featuring highly anisotropic structures in small clusters and a transition toward spherical configurations beyond ~100 \ce{H2O} molecules—suggest specific trends in their infrared (IR) spectra. Smaller clusters are expected to exhibit broader and more structured spectra due to multiple hydration environments, while larger clusters may display a sharper free-OH band that is red-shifted due to enhanced hydrogen bonding. Furthermore, the bonded-OH band is predicted to extend to lower frequencies compared to clusters containing monovalent or divalent cations, reflecting the strong field-induced reorganization of the hydrogen-bond network by \ce{Fe^{3+}}.

One limitation of this study is the lack of explicit electronic polarization 
in the empirical models. However, we addressed this by examining a variety of 
trivalent ions (\ce{Fe^{3+}}, \ce{Lu^{3+}}, \ce{La^{3+}}), all of which exhibited 
similar instability behavior. This consistency suggests that the observed phenomena--particularly 
the shape instability and the positioning of the ion approximately three water 
layers below the surface--are robust and should persist in simulations 
using more advanced models that include electronic polarization.

The charge-induced instability shapes are characterized by conical protrusions on the cluster surface. 
We found that the conical protrusions often host H-bonded pentameric rings.
For the first time, the ability of \ce{H2O} molecules 
to facilitate proton transfer 
in the conical protrusions was examined here. 
To probe the acidity in the conical regions, 
we performed ab initio molecular dynamics (AIMD) simulations on a cluster 
consisting of 64 \ce{H2O} molecules and three \ce{H3O+} ions. Pentameric to
heptameric rings were clearly identified in 
the protonated cluster serving as centers for
proton diffusion. These calculation suggest that within the conical regions, 
protons are able to diffuse across several water molecules, unlike 
the more localized proton delocalization observed in the compact body of the cluster.  
Moreover, AIMD simulations suggest that electronic polarization and charge transfer 
processes delay cluster fragmentation relative to predictions based on empirical force fields.

By analogy, we expect clusters containing trivalent metal ions to exhibit enhanced proton 
delocalization and greater acidity in the conical protrusions at the surface. 
Whether this behavior persists in larger conical protrusions--such as those formed in Rayleigh jets--remains an open question. Notably, the internal electric field present in such jets due 
to their dielectric properties can orient water molecules in ways that potentially 
alter their acid–base characteristics relative to those in the spherical body of the droplet.
We hypothesize that this localized enhancement of acidity may facilitate the charging of analytes and macromolecules entrained within the jet during electrospray ionization. As such, our findings offer insight into the fundamental mechanisms of charge transfer and cluster stability at the nanoscale and may have broader implications for analytical techniques like mass spectrometry.

\begin{acknowledgement}
SC\ is grateful to Prof. D. Frenkel, Yusuf Hamied Department of Chemistry, 
University of Cambridge, UK, Prof.~R. Kapral, Department of Chemistry, The University of Toronto,
Prof. S.S. Xantheas, Pacific Northwest National Laboratory and Dr. Anatoly Malevanets for 
discussions on the stability of charged systems.  SC\ acknowledges an NSERC-Discovery grant (Canada) for funding this research. MPA\ acknowledges a MITACS-Globalink internship fellowship held in the SC group. HN\ acknowledges the province of Ontario 
and The University of Western Ontario for an Ontario Graduate Scholarship. Digital Research Alliance of Canada is acknowledged for providing the computing facilities and technical support to perform this research. 
\end{acknowledgement}

\section*{Supporting Information}

(S1) Models and parameters for MD simulations.  
(S2) System preparation for AIMD. (S3) AIMD Simulation details. 
(S4) Protocol validation for AIMD parameters by testing them in bulk solution simulations.
(S5) Selection of the Poisson solver. (S6) Radial distribution functions between oxygen 
and \ce{Fe^{3+}}, \ce{Lu^{3+}}, and \ce{La^{3+}} in aqueous clusters.
(S7) Distribution of $\cos(\theta)$ around 
\ce{Fe^{3+}}, \ce{Lu^{3+}}, or \ce{La^{3+}} in aqueous clusters. 
(S8) Ring structures in clusters comprising 64~\ce{H2O}-3\ce{H3O+}
ions studied by AIMD.
(S9) Movie that shows the transitions among different ``star-like'' shapes for 
\ce{Fe^{3+}}-$N_{\ce{H2O}}= 78$ at 190~K. (S10) Movie that shows the
proton transfer in $N_{\ce{H2O}}= 64$-3\ce{H3O+} in AIMD simulation.

\begin{suppinfo}
The AIMD data supporting this work are available at the University of 
Western Ontario research data repository,    
Borealis, at https://doi.org/10.5683/SP3/3X3Q4D. 
The data for \ce{La^{3+}}, \ce{Fe^{3+}}, \ce{Lu^{3+}}, 
and pure \ce{H2O} are available at \\ 
https://doi.org/10.7910/DVN/6FGMZS, \\
https://doi.org/10.7910/DVN/MH3AXZ, \\ 
https://doi.org/10.7910/DVN/DHOTEX, \\ 
and https://doi.org/10.7910/DVN/FWB7K2, respectively.
\end{suppinfo}

\section*{Author Contributions}
First authorship is equally shared between JS\ and HN. 
MPA\ contribution: editing of the manuscript,
production and analysis of data, assistance with the usage of software, literature analysis.
SC\ contribution: Conceptualization (100\%),
acquiring funding (100\%), writing and editing of the manuscript,
formal data analysis, literature review and analysis, 
supervision and communication,
handling of submission. 

\providecommand{\latin}[1]{#1}
\makeatletter
\providecommand{\doi}
  {\begingroup\let\do\@makeother\dospecials
  \catcode`\{=1 \catcode`\}=2 \doi@aux}
\providecommand{\doi@aux}[1]{\endgroup\texttt{#1}}
\makeatother
\providecommand*\mcitethebibliography{\thebibliography}
\csname @ifundefined\endcsname{endmcitethebibliography}
  {\let\endmcitethebibliography\endthebibliography}{}

\clearpage


\end{document}


\clearpage

\subsection{S1. Models and parameters for MD simulations}
Molecular dynamics (MD) simulations of aqueous clusters charged by a single \ce{Fe^{3+}}, \ce{Lu^{3+}}, 
and \ce{La^{3+}}, were performed using the Nanoscale Molecular Dynamics\cite{namd} (NAMD) v~2.14 
software and visualized by Visual Molecular Dynamics\cite{HUMP96} (VMD) v~1.9.4. 
The clusters were comprised of $N_{\ce{H2O}}=32-242$~\ce{H2O} molecules. MD simulations were also performed for pristine aqueous clusters with $N_{\ce{H2O}} = 70$ and 242 for using them
as reference systems. For $32<N_{\ce{H2O}}<100$ the cluster sizes were in increments of approximately 
five \ce{H2O} molecules and in particular in the range of $75<N_{\ce{H2O}}<80$ where we were 
searching for the Rayleigh limit, in increments of one molecule.
The water molecules were modelled by the TIP3P-CHARMM  model \cite{mackerell1998all} 
and the ions by the parameters reported in Ref.\cite{won2012force}. 
Rigid three sites water model has been used extensively in the
modeling of \ce{Fe^{2+}} or \ce{Fe^{3+}} ions 
in aqueous systems\cite{guardia1990molecular,kneifel1994simulation,amira2004molecular,li2015parameterization}.   

%

The MD simulations of clusters were performed in the canonical ensemble
by placing the cluster to the center of a spherical cavity where the spherical
boundary condition was applied. The center of mass (COM)
of the cluster was restrained to the center of a spherical cavity.
The restrain of the COM was applied by using the collective variables module, 
COLVARS \cite{fiorin2013using}. The radius of the cavity was set
to three times the radius of the cluster, 
which ensures a sufficient volume to accommodate any large droplet deformations,
such as the formation of conical protrusions.
The electrostatic interactions were treated
with the multi-level summation method\cite{hardy2009multilevel} (MSM) as implemented in NAMD.
In the MSM, a standard cutoff of 12~\AA, with a switch distance of 8~\AA, and a 
pairlistdist of 13.5~\AA were employed. The choice of default grid spacing 2.5~\AA was applied.

The MD runs were performed with the velocity Verlet\cite{velocity-verlet} integrator 
using a time step of 1~fs at 
temperature $T = 90$~K, 130~K, 190~K and 250~K.
The length of the simulation runs was 40~ns.
The temperature of the systems was controlled with the Langevin thermostat
where the damping coefficient in the Langevin
thermostat applied on all atoms was 1~ps.
Among the tested temperatures, $T=190$~K was selected to perform the majority of the simulations.
This temperature was selected because at this temperature 
the molecules in the clusters are still
diffusing within the simulation time.
At $T = 90$~K and 130~K the clusters
demonstrate deformations caused by the instability but there is no internal motion of the 
\ce{H2O} molecules and the ion 
within the simulation time.

\subsection{S2. System preparation for AIMD}
Aqueous clusters containing
$N_{\ce{H2O}}=64$ and 
3\ce{H3O+} ions 
were prepared using the Packmol package\cite{martinez2009packmol} 
and initially simulated with empirical force fields
in the large-scale atomic/molecular massively parallel 
simulator (LAMMPS) package\cite{plimpton1995fast} version Sep 15, 2022. MD simulation was performed in a simulation cell  
with dimensions $48 \times 48 \times 48$ \AA$^3$ using the Nos\'e-Hoover thermostat to set the temperature at 250~K.

In $N_{\ce{H2O}}=64$ the three \ce{H3O+} ions were placed in an isosceles triangle configuration. 
Two of the ions were placed diametrically apart at a distance of 11.2 \AA\, and the third one 
at a distance of 7.92 \AA\ apart from each of the other two \ce{H3O+} ions.

The water molecules were modelled by the SPC/Fw\cite{wu2006flexible} force field. 
The empirical force field \cite{vacha2007autoionization, 10.1063/1.4942771} 
for the \ce{H3O+} ion were taken from Ref.~\cite{10.1063/1.4942771}. 
The MSM\cite{hardy2006multilevel,hardy2009multilevel} was used to treat the 
electrostatic interactions.
In the implementation of MSM in LAMMPS, 
the charged cluster was neutralized by evenly placing 40 
counter ions (with net charge of $-0.0705$) at the outermost circle region 
(with thickness of 10~{\AA}) in the simulation box to compensate the 
system charge of $+3$. 
A command of kspace-style msm/cg $0.35$ was used to set up the cutoff of the atom 
with the net charge below the $=0.35$ would not be considered in the calculation 
of long range electrostatics. 
The cluster fragmented within 3-5~ps of an MD run. Configurations before the fragmentation were collected as starting configuration in a AIMD simulation
in the CP2K software\cite{kuhne2020cp2k} v~9.1.
The details of the CP2K simulations are discussed in the next section. 

\subsection{S3. AIMD Simulation details}
Ab initio MD simulations were performed to study the cluster stability and 
possible proton transfer in these systems. 
All calculations were conducted using the electronic structure module 
Quickstep in the CP2K v~9.1 software package \cite{kuhne2020cp2k}. The 
propagation of the classical nuclei was performed using \textit{ab initio} Born-Oppenheimer
MD with a 0.5~fs timestep. At each step of the MD, the electronic orbitals were
fine-tuned to the Born-Oppenheimer surface using an orbital
transformation method \cite{vandevondele2003efficient} with a 
convergence criterion of $1 \times 10^{-7}$ a.u..
The conjugate gradient minimizer was employed within the orbital transformation 
method to perform the optimization. 
The Gaussian and plane waves method\cite{lippert1999gaussian} expanded the wave function in the Gaussian double-zeta valence polarized (DZVP) basis set \cite{vandevondele2007gaussian}. Auxiliary plane waves were employed to expand the electron density up to a cutoff of 400 Ry. Two density functional approximations, Perdew-Burke-Ernzerhof (PBE) \cite{perdew1996generalized} and Becke-Lee-Yang-Parr (BLYP) \cite{becke1988density,lee1988development}, were used with Grimme dispersion corrections \cite{grimme2006semiempirical} denoted as D3 here. Adopting empirical dispersion correction to density functionals has shown superiority in optimizing water properties at ambient conditions\cite{knight2012multiscale}. Goedecker-Teter-Hutter pseudopotentials were utilized to effectively handle the core electrons \cite{goedecker1996separable}. 
All the protonated clusters were placed in a large vacuum box 
of dimensions $48 \times 48 \times 48$ {\AA}$^3$.
A vacuum shell of 3-\ce5 \AA\ around the clusters was considered to be a sufficient and reasonable range for AIMD simulations of non-periodic isolated systems\cite{marx2009ab}.
The Poisson solver WAVELET\cite{genovese2006efficient,genovese2007efficient} 
was chosen to solve the Poisson equation to satisfy the non-PBC. 

Validation of the AIMD simulation protocol for clusters and to assess the sensitivity
of the proton transfer with respect to the choice of different basis sets
and density functional (PBE-D3-DZVP and BLYP-D3-DZVP), a cubic box containing 64 water molecules with a length of 12.48 \AA\ (corresponding to $1.0 \mathrm{~g} / \mathrm{cm}^3$ density) was
also constructed. 
The validation tests are presented in Sec.~S4 in the Supplementary Information (SI).
The protocol of these reference simulations in the bulk solution was
similar to the clusters, with the only difference being that the system used PBC. 
All AIMD simulations were performed in the canonical ensemble at 300~K with a Nos\'{e}–Hoover 
thermostat (3 chains) at the time constant of 100 fs to regulate the temperature.

The choice of this solver is discussed in Sec.~S5 in SI.

\subsection{S4. Protocol validation for AIMD parameters by testing them in bulk solution simulations}

\begin{figure}[htb!]
\centering
\includegraphics[width=0.65\textwidth]{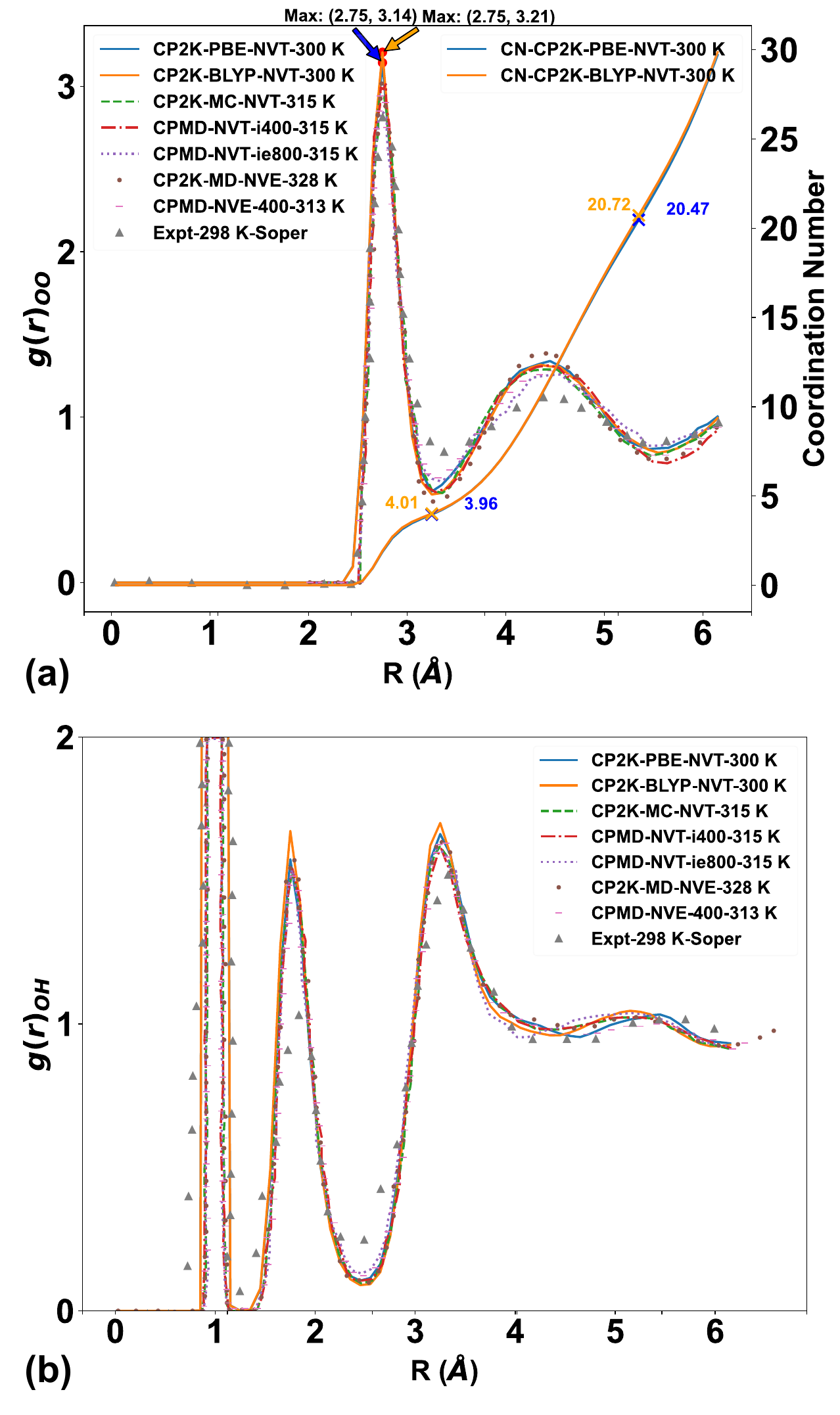}
\caption{Comparison of $g_{OO}$ (a) and $g_{OH}$ (b) derived from simulations and experiment results. The cases `CP2K-PBE-D3-DZVP' and `CP2K-BLYP-D3-DZVP' correspond to the results of our simulations of 64~\ce{H2O} bulk at 300 K. The cases `CP2K-MC-NVT', `CPMD-NVT-i400', and `CPMD-NVT-ie800' are the different AIMD simulations at 315 K by \citeauthor{kuo2004liquid}\cite{kuo2004liquid}. The cases `CP2K-MD-NVE-328 K' and `CPMD-NVE-400-313 k' are also AIMD simulations in NVE ensemble by \citeauthor{kuo2004liquid}\cite{kuo2004liquid}. The cases `PIMD-NVT-300 K' and `ML-Potential-PIMD-NVT-300 K' are the path-integral MD simulation by \citeauthor{chen2016ab}\cite{chen2016ab} and \citeauthor{cheng2019ab}\cite{cheng2019ab}, respectively. The cases `Expt-298 K' correspond the experimental investigations by \citeauthor{chen2016ab}\cite{chen2016ab} and \citeauthor{soper2013radial}\cite{soper2013radial}. The coordination number of water molecules for the $g_{OO}$ results is shown in (a).}
\label{fig:aimd-si-rdf-oo-oh}
\end{figure}

\begin{figure}[htb!]
\centering
\includegraphics[width=0.65\textwidth]{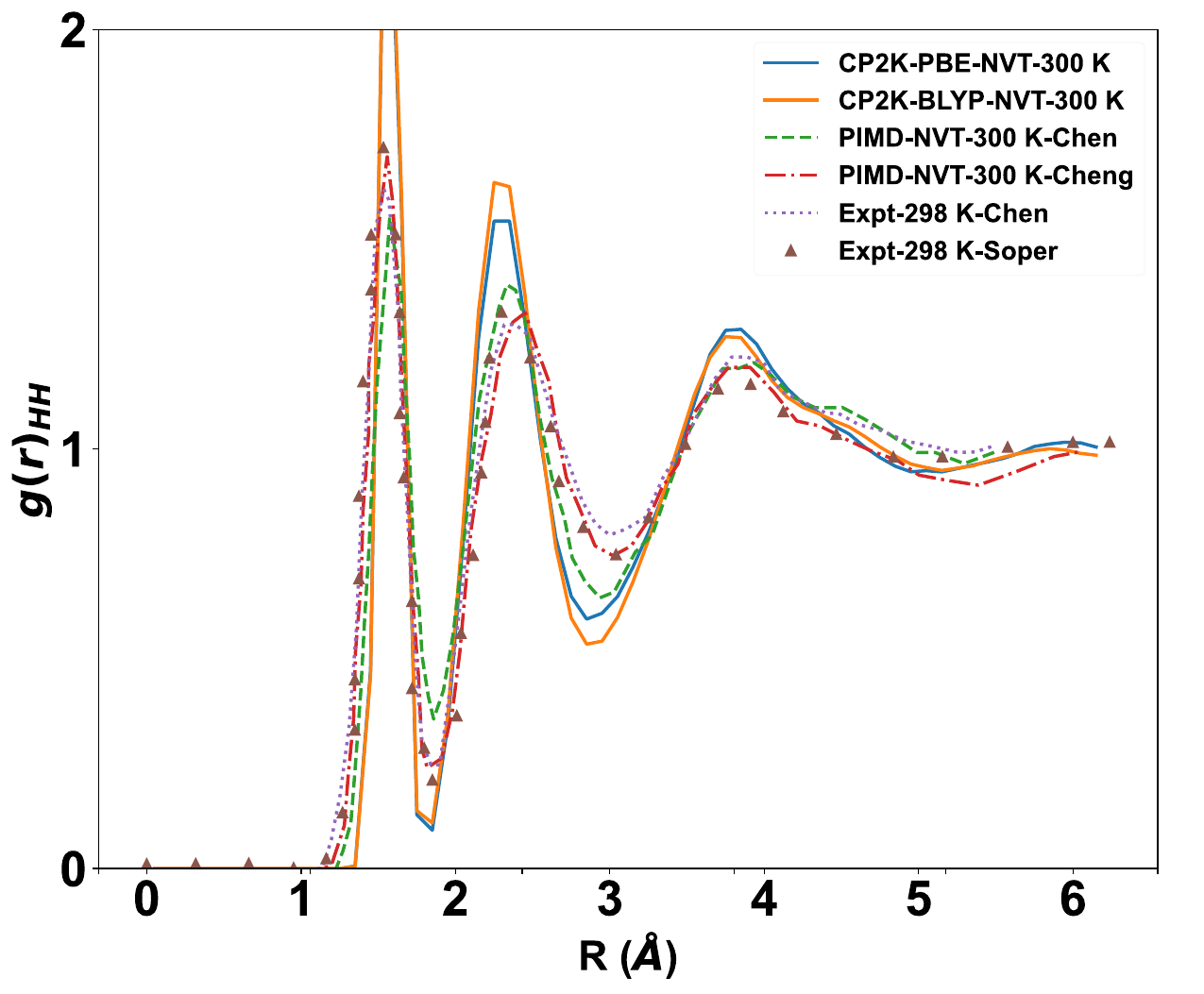}
\caption{Comparison of $g_{HH}$ derived from simulations and experiment results. The cases `CP2K-PBE-D3-DZVP' and `CP2K-BLYP-D3-DZVP' correspond to the results of our simulations of 64~\ce{H2O} bulk at 300 K. The cases `CP2K-MC-NVT', `CPMD-NVT-i400', and `CPMD-NVT-ie800' are the different AIMD simulations at 315 K by \citeauthor{kuo2004liquid}\cite{kuo2004liquid}. The cases `CP2K-MD-NVE-328 K' and `CPMD-NVE-400-313 k' are also AIMD simulations in $NVE$ ensemble by \citeauthor{kuo2004liquid}\cite{kuo2004liquid}. The cases `PIMD-NVT-300 K' and `ML-Potential-PIMD-NVT-300 K' are the path-integral MD simulation by \citeauthor{chen2016ab}\cite{chen2016ab} and \citeauthor{cheng2019ab}\cite{cheng2019ab}, respectively. The cases `Expt-298 K' correspond the experimental investigations by \citeauthor{chen2016ab}\cite{chen2016ab} and \citeauthor{soper2013radial}\cite{soper2013radial}.}
\label{fig:aimd-si-rdf-hh}
\end{figure}

The radial distribution function (RDF) between atoms of water molecules was analyzed to verify the accuracy of our simulation parameters settings and results. The data for our simulated control group was mainly selected from previous work done by \citeauthor{kuo2004liquid} \cite{kuo2004liquid}. The data for the experimental control group was chosen from the work of \citeauthor{soper2013radial} \cite{soper2013radial}, \citeauthor{chen2016ab} \cite{chen2016ab} and \citeauthor{cheng2019ab} \cite{cheng2019ab}, respectively. All the details of the simulations and experiments in the control group are labelled in Fig.~\ref{fig:aimd-si-rdf-oo-oh} and Fig.~\ref{fig:aimd-si-rdf-hh} and summarized in Table~\ref{tbl:rdf-system} for a more precise presentation. For the RDF between the oxygen atoms of the water molecules, it can be found that the simulation and experimental results agree at the first neighborhood of the water molecules, at 2.75 \AA. However, there are some discrepancies in the peaks, which range from 2.85 to 3.21, with the mean of the data at 
$3.01 \pm 0.13$. The relevant data are included in Table~\ref{tbl:rdf-goo}. 
The results of our two simulations with different combinations are similar. 
We also calculated the coordination number (CN) for the water molecules first and second nearest neighbours. 
The CNs were determined by calculating the oxygen-oxygen number integral at the location of the first 
minimum in the associated RDF. We can find that in the first shell, the CNs of the water molecules are 
very near to 4. All the simulation algorithms from \citeauthor{kuo2004liquid}'s and other experimental results show a high degree of agreement in the first peak of the RDFs, with Monte Carlo calculations using CP2K \cite{kuhne2020cp2k} in the NVT ensemble at 315 K, simulation using Car-Parrinello Molecular Dynamics (CPMD) code \cite{Hutter2005CPMDCM} in the NVT ensemble at 315~K (with the value of the fictitious electron mass at $\mu = 400$ a.u.), and simulation using CPMD in the NVE ensemble at 313 K. Moreover, there is good correspondence with our work at the second peak. However, one of the simulations using CPMD \cite{Hutter2005CPMDCM} at 315~K in the NVT ensemble (with the more significant value of the fictitious electron mass at $\mu = 800$ a.u.) and one 
using CP2K \cite{kuhne2020cp2k} in the NVE ensemble at 328 K, and the experimental results
of \citeauthor{soper2013radial}'s show some minor deviations from our results in the second peak. 
Yet \citeauthor{kuo2004liquid} 
argued using a $\mu = 800$ a.u value ensures stable CPMD trajectories for simulations
in the canonical ensemble while employing large thermostats of the nuclear and electronic degrees of freedom.

\begin{table*}
\caption{Details of the setup of the simulations used to compare them with each other in Fig.~\ref{fig:aimd-si-rdf-oo-oh} and Fig.~\ref{fig:aimd-si-rdf-hh}. MC denotes Monte Carlo simulations. The designations `i' and `ie' represent the use of massive thermostats specifically for the ionic degrees of freedom or both ionic and electronic degrees of freedom, respectively. Some simulations used two distinct options for the fictitious electron mass: a small value of $\mu = 400$ a.u. and a larger value of $\mu = 800$ a.u. $\Delta t$ and $t_{prod}$ are the timestep and length of production run \cite{kuo2004liquid}.}
\label{tbl:rdf-system}
\resizebox{\textwidth}{25mm}{
\begin{tabular}{cccccccccc}
\hline
&functional&basis set&pseudopotential&cut-off [Ry]&$\mu$ [a.u.]&N&$\Delta$ t [fs]&$t_{prod}$ [ps]&T [K] \\
\hline
CP2K-MD-NVT&PBE-D3&DZVP&GTH&400&n/a&64&0.5&5&300\\
CP2K-MD-NVT&BLYP-D3&DZVP&GTH&400&n/a&64&0.5&5&300\\
CP2K-MC-NVT&BLYP&TZV2P&GTH&280&n/a&64&n/a&n/a&315\\
CPMD-NVT-i400&BLYP&PW&TM&85&400&64&0.097&11.1&315\\
CPMD-NVT-ie800&BLYP&PW&TM&85&800&64&0.097&7.2&315\\
CP2K-MD-NVE&BLYP&TZV2P&GTH&280&n/a&64&0.5&12.0&328\\
CPMD-NVE-400&BLYP&PW&TM&85&n/a&64&0.48&11.4&328\\
PIMD-NVT&PBE-rVV10&PW&norm-conserving&85&n/a&32&0.5&8.0&300\\
PIMD-NVT&revPBE0-D3&TZV2P&GTH&400&n/a&64&0.5&100&300\\
\hline
\end{tabular}}
\end{table*}

\begin{table*}
\caption{Comparison of $g_{OO}$ simulation and experiment results in Fig. \ref{fig:aimd-si-rdf-oo-oh}(a). The values of the height and location of the first peak in the oxygen-oxygen RDFs, as well as the oxygen-oxygen CNs are presented.}
\label{tbl:rdf-goo}
\resizebox{\textwidth}{20mm}{
\begin{tabular}{ccccccc}
\hline
&functional&basis set&pseudopotential&$r_{OO}$,max [\AA]&$g_{OO}$,max&CN \\
\hline
CP2K-MD-NVT&PBE&DZVP&GTH&2.75&3.14&3.96\\
CP2K-MD-NVT&BLYP-D3&DZVP&GTH&2.75&3.21&4.01\\
CP2K-MC-NVT&BLYP&TZV2P&GTH&2.75&3.0&4.0 \\
CPMD-NVT-i400&BLYP&PW&TM&2.75&3.0&4.0 \\
CPMD-NVT-ie800&BLYP&PW&TM&2.75&2.9&4.0 \\
CP2K-MD-NVE&BLYP&TZV2P&GTH&2.75&3.1&4.0 \\
CPMD-NVE-400&BLYP&PW&TM&2.75&2.9&4.0 \\
Experiment&n/a&n/a&n/a&2.75&2.85&4.0 \\
\hline
\end{tabular}}
\end{table*}

The RDFs of oxygen-hydrogen and hydrogen-hydrogen atom pairs were also calculated. Our results for the RDFs between hydrogen and oxygen atoms perfectly agree with the experimental results at the first peak. They differ somewhat from the other simulations, but all the results generally have a similar trend. Our simulation results vary slightly from all other experiments and simulations for the RDFs between hydrogen and hydrogen atoms but still follow the same trend. These results are shown in Fig. \ref{fig:aimd-si-rdf-hh}. By calculating the RDFs between different atoms of water molecules in the 64-water bulk system and comparing them with other typical simulation and experimental results, we can see that our results are accurate and the simulation parameters are set up correctly. The simulation results are similar by combining two different basis sets and density functional approximations, PBE-D3-DZVP and BLYP-D3-DZVP.

\clearpage

\subsection{S5. Selection of the Poisson solver}
The choice of Poisson solver can significantly impact the accuracy and reliability of the results when computational simulations involving isolated systems. Different solvers were designed to solve the Poisson equation,
\begin{equation}
\nabla^2 \phi(\mathbf{r})=-\frac{\rho(\mathbf{r})}{\epsilon_0}
\end{equation}
where $\nabla^2$ is the Laplacian operator, $\phi(\mathbf{r})$ is the electrostatic potential, 
$\rho(\mathbf{r})$ is the charge density at position $\mathbf{r}$, 
$\epsilon_0$ is the permittivity of free space. The Poisson solvers handle the electrostatic potential 
and boundary conditions in distinct ways, which can lead to variations in the computed properties of the system. 
To understand these differences, comparing and discussing the simulation results produced 
by different Poisson solvers is essential.
In this work, we focus on two specific Poisson solvers: 
the WAVELET \cite{genovese2006efficient,genovese2007efficient} and the 
MT (Martyna-Tuckerman) \cite{martyna1999reciprocal}. 
 
One of the significant advantages of the wavelet-based Poisson solver 
is its flexibility in handling boundary conditions. Unlike traditional methods that may require artificial periodicity or complex boundary condition treatments, wavelets naturally accommodate a variety of boundary conditions, including Dirichlet \cite{evans2022partial}, Neumann \cite{evans2022partial}, or mixed boundary conditions. This makes the method well-suited for isolated systems or systems with complex geometries. Hence, we prefer the WAVELET solver for calculating our neutral and charged clusters. The MT \cite{martyna1999reciprocal} Poisson solver is another advanced method implemented in CP2K to address the issues of simulating systems with partial periodicity within a computational framework that generally uses PBC. It is useful for simulations where we need to minimize or eliminate the artificial interactions between periodic images, making it ideal for systems like slabs, surfaces, or even clusters that are non-periodic in at least one dimension. 

\begin{figure*}[htbp!] 
\centering    
\includegraphics[width=1.0\textwidth]{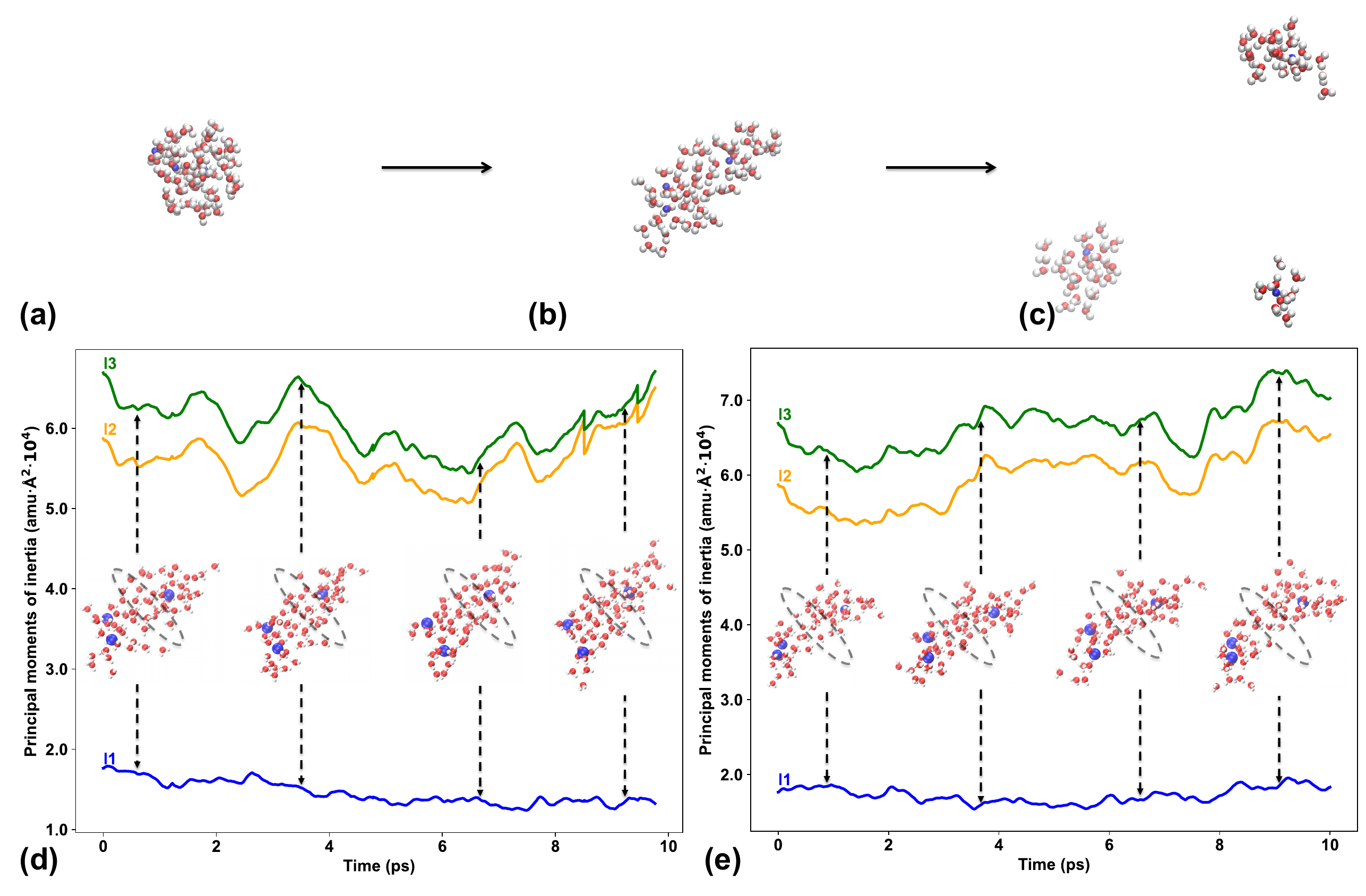}
\caption{(a) The initial configuration of the 64-water cluster containing three \ce{Na+}. (b) A typical intermediate configuration of the 64-\ce{H2O}-3\ce{Na+} cluster before separating into three sub-clusters. This configuration was used as the initial configuration for performing AIMD simulation in CP2K. (c) The final configuration of the 64-\ce{H2O}-3\ce{Na+} cluster separates into three sub-clusters after 1~ns MD simulation in the $NVT$ ensemble. Configuration (b) is used to initiate a CP2K run up to 10~ps using the WAVELET Poisson solver for which the configuration after 10~ps is shown in (d) and using the MT Poisson solver shown in (e). The principal moments of inertia reflect the extension and contraction of the cluster in different directions, respectively. As the simulation progresses, the elliptical dashed line in (d) and (e) marks snapshots of the simulation that show a progressively narrower bottleneck in the junction, reflecting a tendency for the entire cluster to divide into two
sub-clusters.}
\label{sodium}
\end{figure*}

Analyzing the simulation results using these two approaches, we can gain insight into their respective strengths and limitations and their suitability for different isolated systems. This comparison will help elucidate how the choice of Poisson solver influences the physical observables of interest and contribute to a more informed selection of computational methods in further studies. Before we finally chose which Poisson solver to use in the 64-water cluster containing three hydroniums, we prepared a 64-water cluster containing three \ce{Na+} ions and 
counter ions in the $80\times 80\times 80$ cubic cell to make a similar system, 
as shown in Fig.~\ref{sodium}~a. We performed MD simulations up to 1 ns for this 
system in the NVT ensemble using the LAMMPS (21 Nov 2023 version) 
package \cite{thompson2022lammps}. We used the Multi-Level Summation Method 
(MSM) solver \cite{hardy2006multilevel} to compute long-range interactions in MD simulations 
with non-PBC. The entire cluster was placed in the centre of a cubic vacuum box 
with dimensions $50 \times 50 \times 50 \AA^3$. Then the counterions were placed close to the edge of 
the cubic box to ensure that the interaction force of the counterions on the clusters was negligible. 
The final configuration of the MD simulation, as shown in Fig.~\ref{sodium}~(c), presents 
the division of the whole cluster into three sub-clusters, each containing a 
solvated \ce{Na+} ion. Fig.~\ref{sodium}~(b) shows the configuration before the cluster separates, 
which tends to separate as shown by its elongation but is still connected. 
We then used the configuration from Fig.~\ref{sodium}~(b) as our initial structure 
in CP2K and continued with the AIMD simulation. We applied WAVELET and MT Poisson 
solvers in this simulation to handle the electrostatic forces in a non-PBC setting. 
After 10~ps throughout the simulation, we observe a tendency for one of the 
\ce{Na+} ions to separate from the cluster in both simulations, as shown in the snapshots of the 
simulations in Figs.~\ref{sodium}~(d) and Fig.~\ref{sodium}~(e), respectively. 
The whole cluster continues to elongate as the simulation proceeds, and the grey 
dashed circles mark that the cluster becomes gradually narrower in the interior, with a tendency to divide. 
This conformational elongation and deformation process is also clearly monitored by the
principal moments of inertia parameter. The main axis I1 in Fig.~\ref{sodium}~(d) and
Fig.~\ref{sodium}~(e) are through the cluster and orientated parallel to its elongation direction. 

Comparing these two figures, we can see that the simulation results using WAVELET in Fig.~\ref{sodium}~(d) more logically illustrate that the principal moments of inertia for this axis I1 show a decreasing trend, corresponding to the other two parameters of the axis I2 and I3 that go through the cluster but are oriented perpendicularly to the axis I1 show an increasing trend. This reflects a process whereby the cluster is elongated, 
and more molecules fit more closely to the I1 axis and are distributed along the elongated direction. Hence, we will use this WAVELET Poisson solver to handle the charged clusters in our main work.

\clearpage

\subsection{S6. Radial distribution functions between oxygen and \ce{Fe^{3+}}, \ce{Lu^{3+}}, and \ce{La^{3+}} in aqueous clusters}

\begin{table*}[htbp!]
    \centering
    \label{table:rdf_o_ion}
    \resizebox{1.0\textwidth}{2cm}{
    \caption{Location of the maxima and coordination number (CN) of the RDF
    between a trivalent ion and oxygen sites of TIP3P water in $N_{\ce{H2O}}=78$ at 190~K. The location of the maxima in
    the RDF peaks in each solvation shell is in {\AA}. 
    $\mathrm{CN}^{(n)}$ ($n = 1-4$) denotes the coordination number in each solvation shell.}
    \begin{tabular}
    {|c|c|c|c|c|c|c|c|c|}
        \hline
        Ion & 1st max & $\mathrm{CN}^{(1)}$ &  2nd max  & $\mathrm{CN}^{(2)}$ & 3rd max & $\mathrm{CN}^{(3)}$ & 4th max & $\mathrm{CN}^{(4)}$ \\
        \hline
        \ce{Fe^{3+}} & 1.6 & 3.5 & 2.9 & 7 & 4.1 & 16 & 5.3 & 53 \\
        \hline
        \ce{Lu^{3+}} & 2.3 & 7 & 3.5 & 9 & 4.5 & 23 & 6.3 & 52\\
        \hline
        \ce{La^{3+}} & 2.6 & 9 & 4.9 & 28 & 6.6 & 54  & & \\
        \hline
    \end{tabular}}
\end{table*}

\begin{figure}[htbp!]
\includegraphics[width=\textwidth]{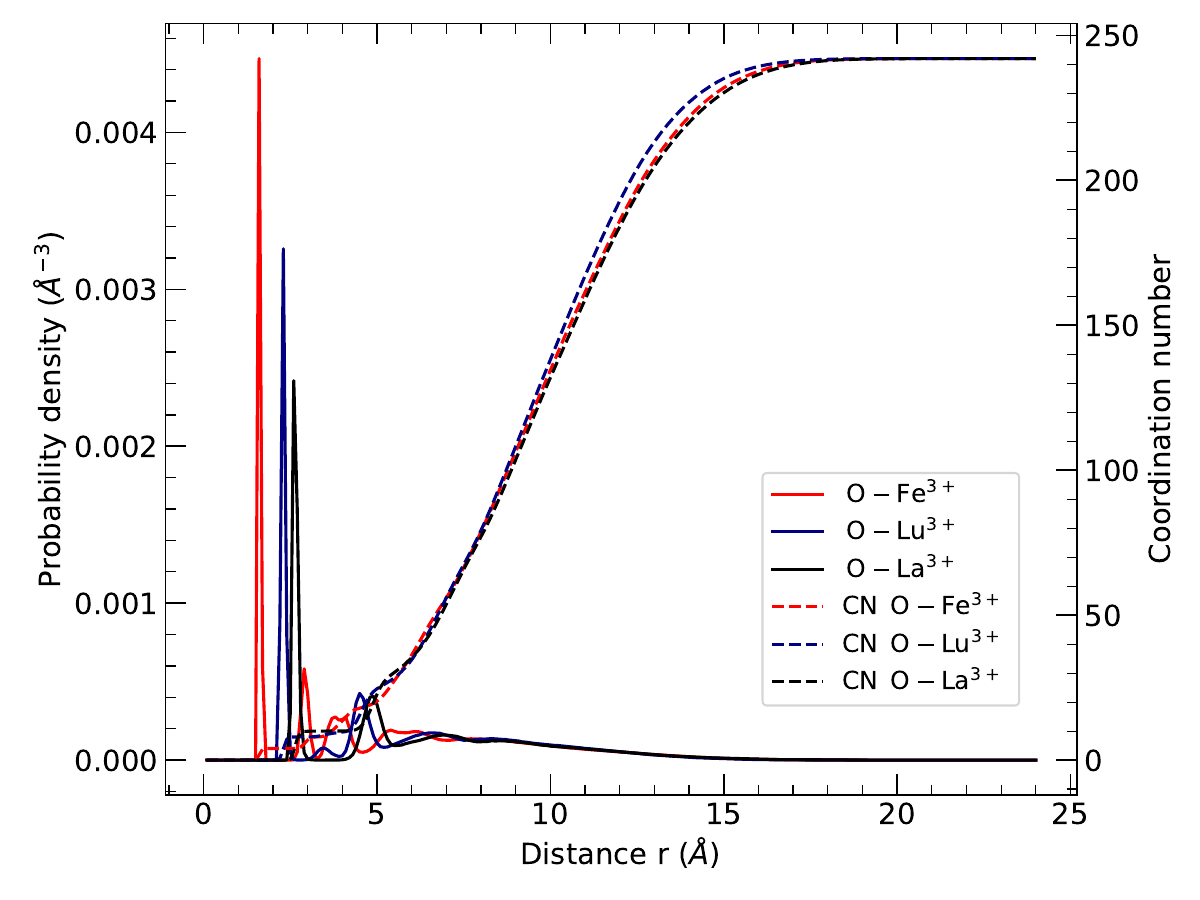}
\caption{Radial distribution function (RDF) and coordination number between oxygen sites of \ce{H2O} and \ce{Fe^{3+}} ion (red colored line), 
\ce{Lu^{3+}}-O (blue colored line), and \ce{La^{3+}}-O (black colored line) in $N_{\ce{H2O}} = 242$ at $T=190$~K.} 
\label{fig:rdf-Oh2-N242-trivalentions}
\end{figure}

\begin{figure}[htbp!]
\includegraphics[width=\textwidth]{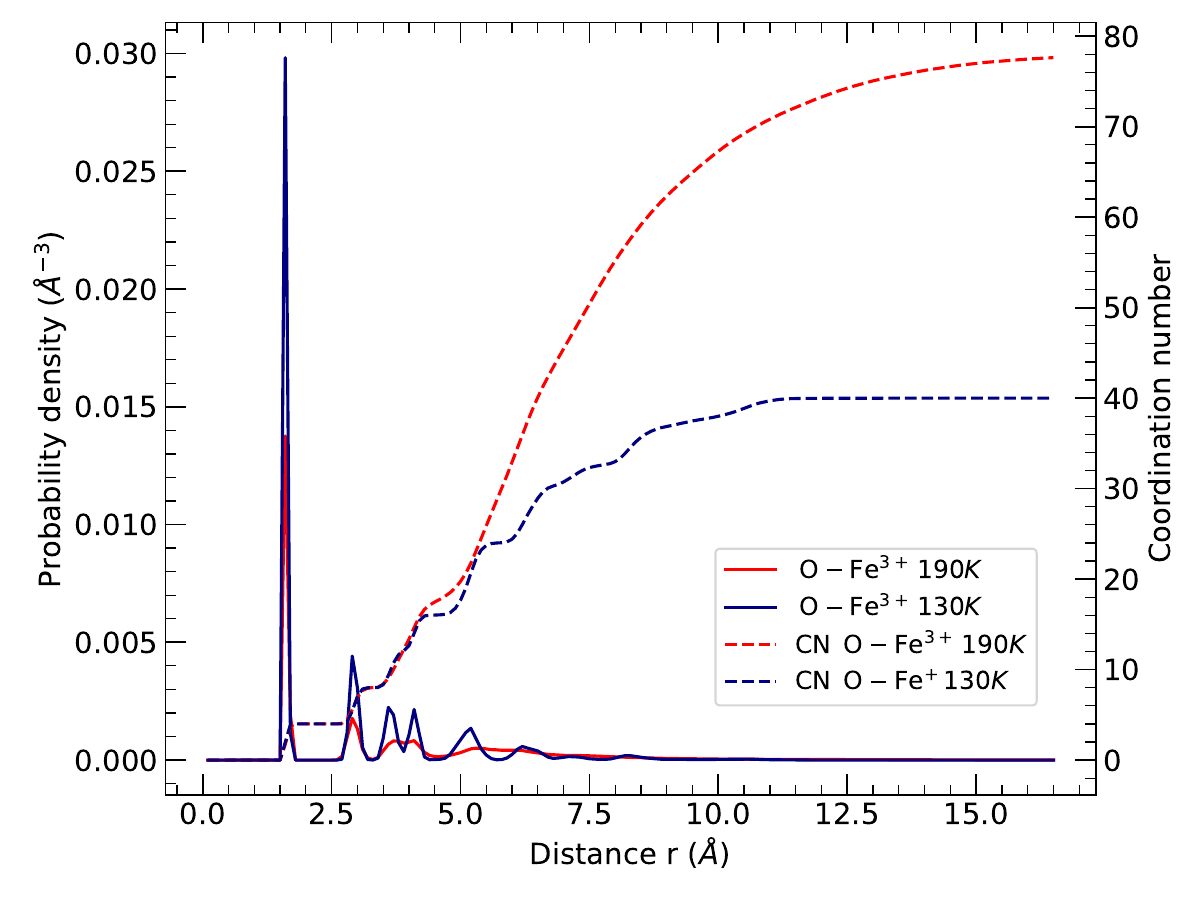}
\caption{RDF and coordination number for O-\ce{Fe^{3+}}  when  $N_{\ce{H2O}} = 40$ 
at $T=130$~K compared with O-\ce{Fe^{3+}}  when  $N_{\ce{H2O}} = 78$ at $T=190$~K.} 
\label{fig:rdf-Oh2-trivalentions-130K}
\end{figure}

\clearpage

\subsection{S7. Distribution of $\cos(\theta)$ around \ce{Fe^{3+}}, \ce{Lu^{3+}}, or \ce{La^{3+}} in aqueous clusters}
\begin{figure}[htbp!]
\centering
\begin{subfigure}[htbp]{0.5\textwidth}
\centering
\includegraphics[width=0.92\textwidth]{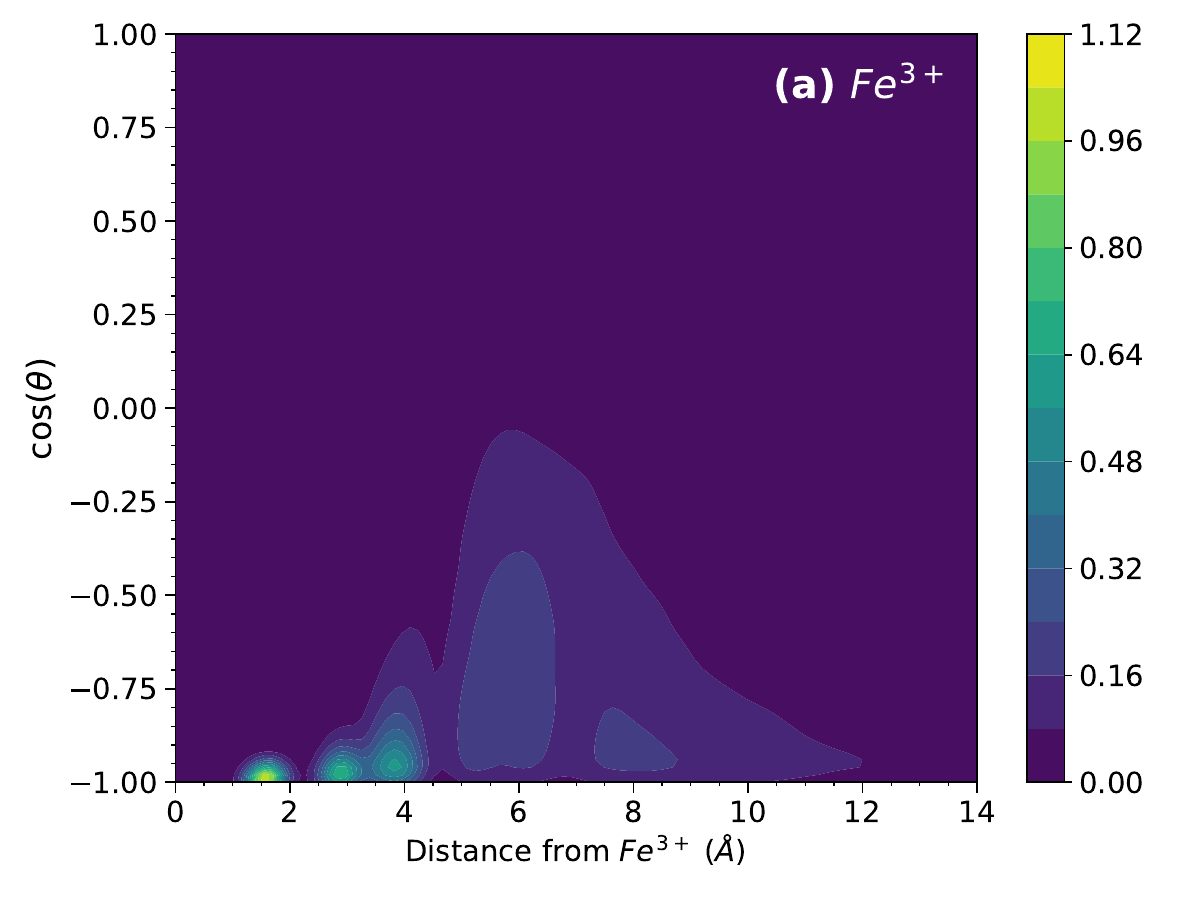}
\caption{}
\end{subfigure}
\begin{subfigure}[htbp]{0.5\textwidth}
\centering
\includegraphics[width=.92\textwidth]{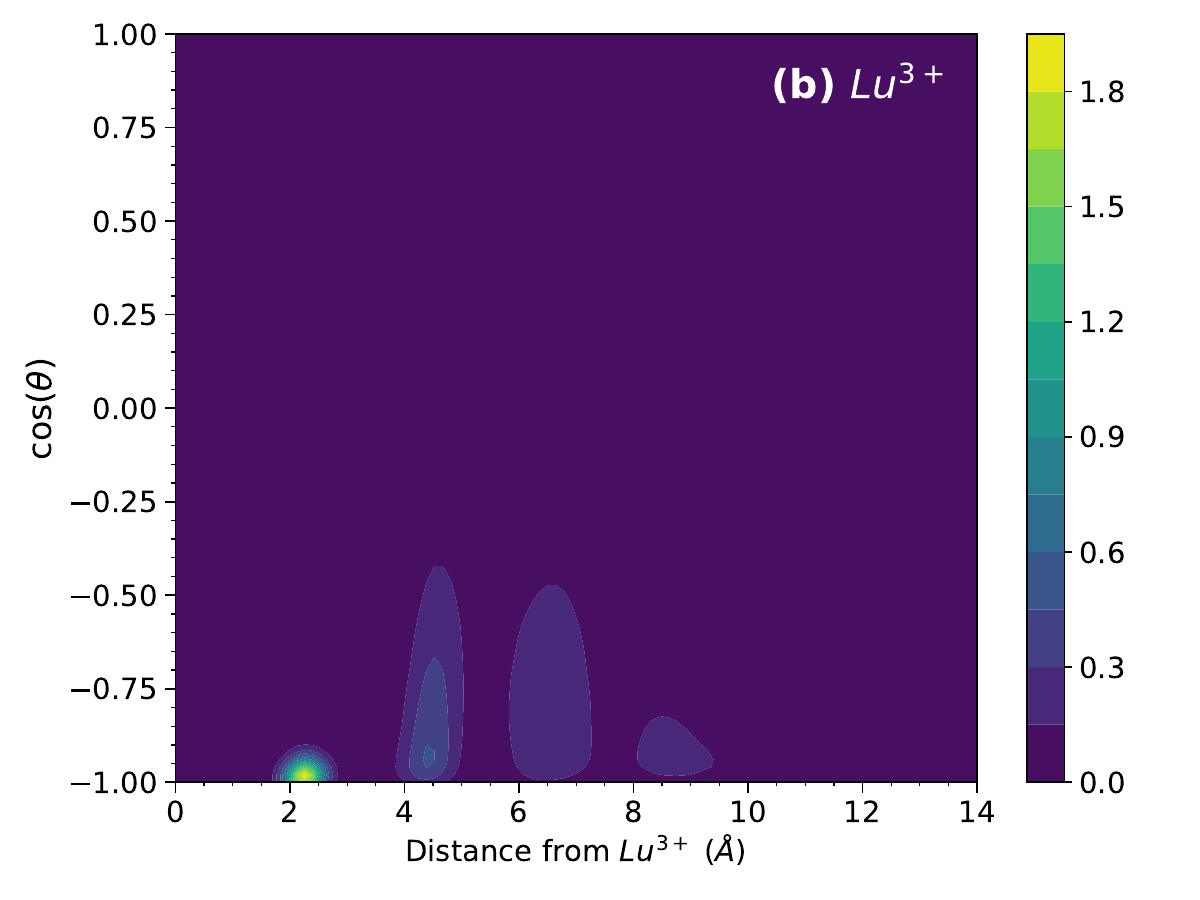}
\caption{}
\end{subfigure}
\begin{subfigure}[htbp]{0.5\textwidth}
\centering
\includegraphics[width=0.92\textwidth]{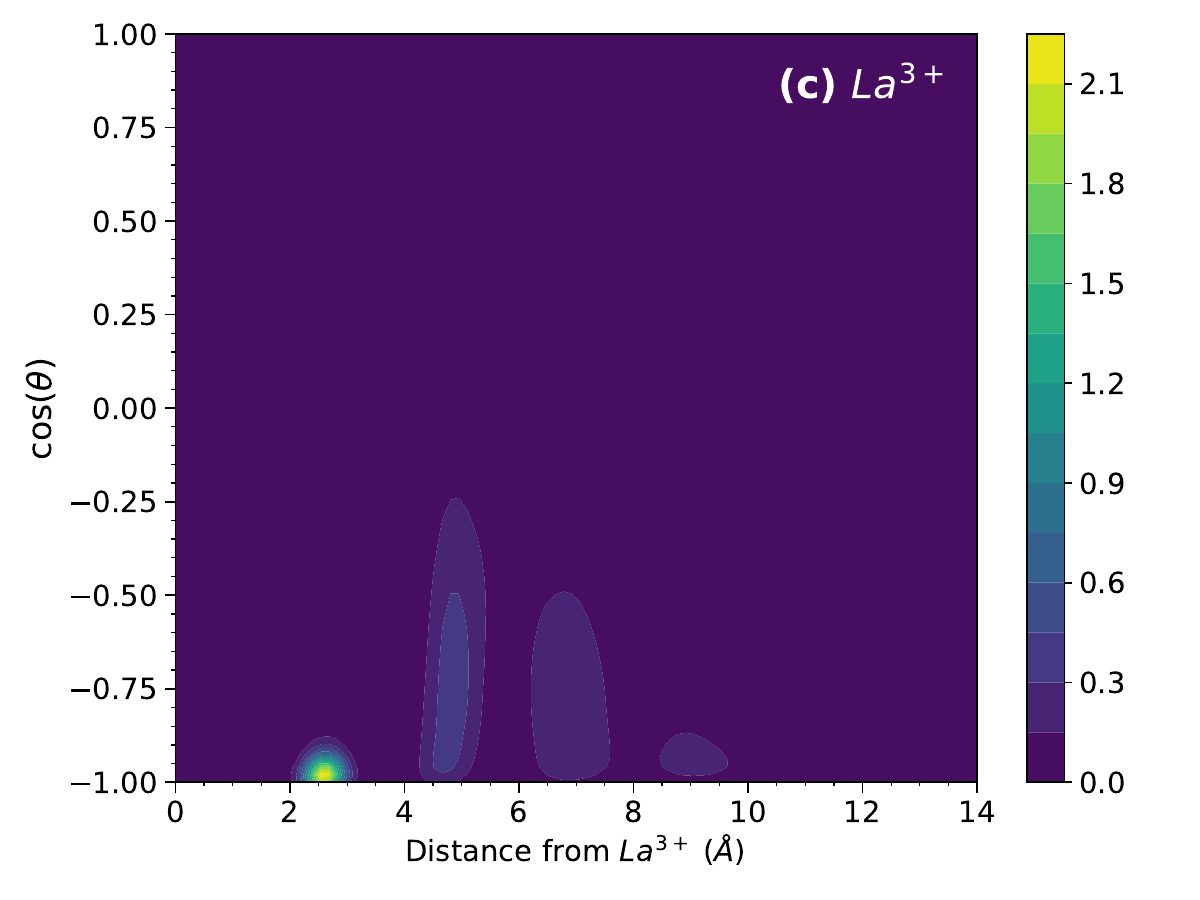}
\caption{}
\end{subfigure}
\caption{Contour maps of $\cos(\theta)$ distribution of \ce{H2O} dipole orientations 
within successive spherical shells centered at (a) \ce{Fe^{3+}}, (b) \ce{Lu^{3+}}, and (c) \ce{La^{3+}} 
in $N_{\ce{H2O}} = 78$ at 190~K.}
    \label{fig:costheta-dipoleMoment-78h2o-190K}
\end{figure}

\clearpage

\subsection{S8. Ring structures in clusters comprising 64~\ce{H2O}-3\ce{H3O+}
ions studied by AIMD}

\begin{figure}[htbp!]
\centering
\includegraphics[width=0.47\textwidth]{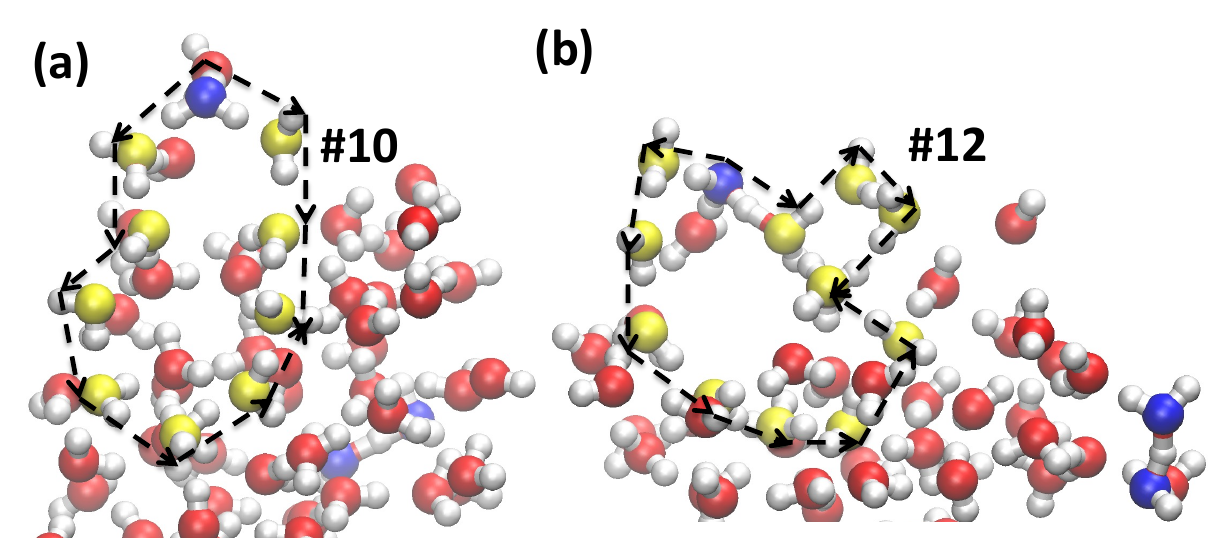}
\caption{Snapshots of the two longer ring structures composing of (a) 10 and (b) 12 \ce{H2O}
molecules from the AIMD trajectory of the 64~\ce{H2O}~-~3~\ce{H3O+} cluster.
These rings appear with low probability in the 7~ps AIMD run.
The color coding of the atomic sites is the same as that in Fig.~8 in the main text. Here, the oxygen sites in the rings are highlighted in yellow. The paths for proton transfer along the rings are indicated by using the dashed arrows.}
\label{c64-3h3o-ring-examples}
\end{figure}

\newpage
\providecommand{\latin}[1]{#1}
\makeatletter
\providecommand{\doi}
  {\begingroup\let\do\@makeother\dospecials
  \catcode`\{=1 \catcode`\}=2 \doi@aux}
\providecommand{\doi@aux}[1]{\endgroup\texttt{#1}}
\makeatother
\providecommand*\mcitethebibliography{\thebibliography}
\csname @ifundefined\endcsname{endmcitethebibliography}
  {\let\endmcitethebibliography\endthebibliography}{}